\documentclass[amsmath,superscriptaddress,showpacs,prl,twocolumn]{revtex4-1}
\usepackage{bm}
\usepackage{enumerate}
\usepackage{graphicx}
\usepackage[dvips]{epsfig}
\usepackage{epsf}

\makeatletter
\newcommand{\Rmnum}[1]{\expandafter\@slowromancap\romannumeral #1@}
\makeatletter

\begin{document}
\title{Magnon spin Nernst effect in antiferromagnets}
\author{Vladimir A. Zyuzin}
\affiliation{Department of Physics and Astronomy and Nebraska Center for Materials and Nanoscience, University of Nebraska, Lincoln, Nebraska 68588, USA}
\author{Alexey A. Kovalev}
\affiliation{Department of Physics and Astronomy and Nebraska Center for Materials and Nanoscience, University of Nebraska, Lincoln, Nebraska 68588, USA}
\begin{abstract}
We predict that a temperature gradient can induce a magnon-mediated spin Hall response in an antiferromagnet with non-trivial magnon Berry curvature.
We develop a linear response theory which gives a general condition for a Hall current to be well defined, even when the thermal Hall response is forbidden by symmetry. We apply our theory to a honeycomb lattice antiferromagnet and discuss a role of magnon edge states in a finite geometry.
\end{abstract}
\maketitle

\noindent
Understanding spin transport in nanostructures is a long-standing problem in the field of spintronics \cite{Dyakonov, Zutic:RoMP2004,Bader.Parkin:ARoCMP2010}. The discovery of the spin Hall effect \cite{Dyakonov.Perel:PLA1971,Hirsch:PRL1999,Zhang:PRL2000, Murakami:S2003,Sinova:PRL2004,Kato.Myers.ea:S2004,Valenzuela.Tinkham:N2006} has been extremely important as it has led to many important developments in spintronics \cite{Sinova.Valenzuela.ea:RoMP2015}, such as the quantum spin Hall effect \cite{Kane.Mele:PRL2005,Bernevig.Hughes.ea:S2006}, the spin-orbit torque \cite{Miron.Garello.ea:N2011,Liu.Lee.ea:PRL2012,Liu:Science2012}, and the spin Seebeck effect \cite{Uchida:Nature2008,Uchida.Xiao.ea:NM2010,Jaworski.Yang.ea:NM2010}. In the instrinsic spin Hall effect, the time reversal symmetry prohibits the transverse charge current but allows the transverse spin current originating from the non-trivial Berry curvature of electron bands \cite{Murakami:S2003,Sinova:PRL2004}. The quantization of the intrinsic spin Hall effect can be characterized by the topological Chern number and is accompanied by the existence of topologically protected edges in the finite geometry \cite{Qi.Wu.ea:PRB2006}. On the other hand the quantum spin Hall effect can be characterized by the Z2 topological invariant \cite{Kane.Mele:PRL2005,Bernevig.Hughes.ea:S2006}.     

The thermal Hall effect carried by magnons has been experimentally observed in collinear ferromagnets such as $\mathrm{Lu}_{2}\mathrm{V}_{2}\mathrm{O}_{7}$, 
$\mathrm{Ho}_{2}\mathrm{V}_{2}\mathrm{O}_{7}$, and 
$\mathrm{In}_{2}\mathrm{Mn}_{2}\mathrm{O}_{7}$ with pyrochlore structure \cite{Onose.Ideue.ea:S2010,Ideue.Onose.ea:PRB2012}. It has been understood that the Dzyaloshinskii-Moriya interaction (DMI) leads to  the Berry curvature of magnon bands and to the transverse with respect to the external temperature gradient energy current \cite{Katsura.Nagaosa.ea:PRL2010,Matsumoto.Murakami:PRL2011,
Zhang.Ren.ea:PRB2013,Lee.Han.ea:PRB2015}. The same effect has also been observed in kagome ferromagnet $\mathrm{Cu}(1-3, \mathrm{bdc})$ \cite{Hirschberger.Chisnell.ea:PRL2015}. The existence of magnon edge states and tunable topology of magnon bands have been discussed theoretically \cite{Matsumoto.Murakami:PRL2011, Shindou.Matsumoto.ea:PRB2013,Zhang.Ren.ea:PRB2013,Shindou.Ohe.ea:PRB2013,Mook.Henk.ea:PRB2014,Mook.Henk.ea:PRB2014a}.  The spin Nernst effect (SNE) has been theoretically studied in Ref.~\cite{Kovalev.Zyuzin:PRB2016} for a kagome lattice ferromagnet. Topological properties of honeycomb lattice ferromagnet were addressed in Refs.~\cite{Fransson.ea:PRB16,Owerre:apa2016,Kim.Ochoa.ea:apa2016}. 

It has been recently realized that antiferromagnets are promising materials for spintronics applications \cite{Jungwirth.Marti.ea:NN2016}.  In Refs.~\cite{Ohnuma.Adachi.ea:PRB2013,Rezende.Rodriguez-Suarez.ea:PRB2016} the spin Seebeck effect has been studied in antiferromagnets.
In Ref.~\cite{Matsumoto.Shindou.ea:PRB2014} it has been shown that the Berry curvature can result in non-zero thermal Hall effect carried by magnons in magnets with dipolar interaction and in antiferromagnets.
However, SNE in antiferromagnets has not been addressed as all of the studies of anomalous magnon-mediated spin transport in magnetic materials have so far been done in ferromagnetic systems. 
 
\begin{figure} \centerline{\includegraphics[clip,width=1\columnwidth]{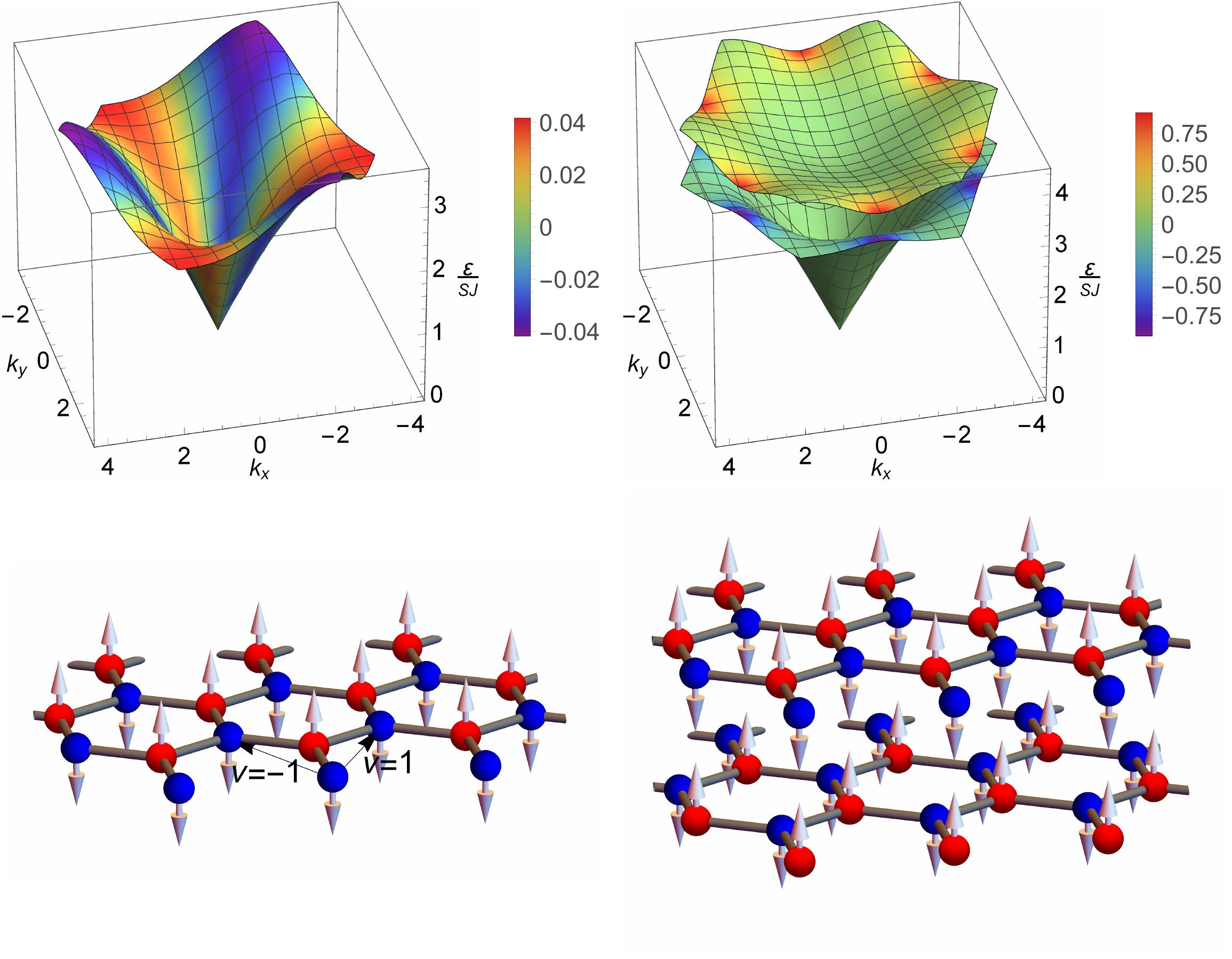}}

\protect\caption{(Color online) 
Left: Magnon spectrum of a single layer antiferromagnet with DMI $D=0.1J$ (black arrows correspond to $\nu$ sign convention of DMI), with schematics of the lattice and order in $z-$direction in the bottom.    
Right: Magnon spectrum of antiferromagnet on a bilayer honeycomb lattice. Parameters are chosen to be $J^{\prime} = J$ and $D=0.1J$.
In both cases the distribution of the Berry curvature over the Brillouin zone is plotted by the color distribution on top of the spectrum for one of the degenerate subbands.   }

\label{fig:honeycomb}  

\end{figure}

In this paper, we study SNE in antiferromagnets with Neel order. 
We first derive a general operator that has a well defined current in a general antiferromagnet. 
We then develop a linear response theory for such a current using the Luttinger approach of the gravitational
scalar potential \cite{Luttinger:PR1964,Tatara:PRB2015}.
It is shown that the response is driven by a modified Berry curvature of magnon bands. We then apply our findings to antiferromagnets with Neel order where a well defined current corresponds to the spin density.
Various realizations of antiferromagnets with honeycomb arrangement of magnetic atoms have been suggested recently \cite{Tsirlin.Janson.ea:PRB2010,Liu.Berlijn.ea:PRB2011,
Lee.Choi.ea:JoPCM2012,Singh.Manni.ea:PRL2012,Choi.Coldea.ea:PRL2012}.
We consider  a single- and bi-layer honeycomb antiferromagnets with antiferromagnetic interlayer coupling where the nearest neighbor exchange interactions and the second nearest neighbor DMI are present (see Fig.~\ref{fig:honeycomb}).  We show that both models possess the magnon edge states in the finite geometry and discuss their role for SNE. For a single layer, we observe an interplay between the Berry curvature due to the lattice topology and DMI and find that the Berry curvature is not of the monopole type, contrary to a ferromagnet on a honeycomb lattice \cite{Kim.Ochoa.ea:apa2016,Owerre:apa2016}. We also find that SNE can be present in antiferromagnets that are invariant under (i) a global time reversal symmetry (e.g. Fig.~\ref{fig:honeycomb}, right) or under (ii) a combined operation of time reversal and inversion symmetries (e.g. Fig.~\ref{fig:honeycomb}, left) which prohibits the thermal Hall response derived in \cite{Matsumoto.Shindou.ea:PRB2014}. 

\noindent
\textit{\underline{Current in antiferromagnet}.}
Here we assume a  general model of antiferromagnet insulator with a magnetic unit cell having $N$ sites.
The Hamiltonian of such a system is of Heisenberg type with exchange interactions, DMI, anisotropies and others. Assuming that we know the order of the system, we study the magnon excitations around that order. The Holstein-Primakoff transformation from spins to boson operators can be employed to study the magnons (see  \cite{Auerbach:GTiCP1994} for example). In this way, the boson operators $\nu_{j}({\bf r})$ and $\nu^{\dag}_{j}({\bf r})$, with $j \in (1,N)$, correspond to $j$th element of the magnetic unit cell.  The operators satisfy commutation relationship $[\nu_{i}({\bf r}),\nu^{\dag}_{j}({\bf r}^{\prime}) ] = \delta_{ij}\delta_{{\bf r}{\bf r}^\prime}$.
We then proceed to write a general form of a Hamiltonian describing the magnons,
\begin{align}\label{hamaf}
H_{0} = \frac{1}{2} \int d{\bf r}\Psi^{\dag}({\bf r}){\hat H}\Psi(\bf r).
\end{align}
Since this Hamiltonian describes magnons of an antiferromagnet, it will necessary contain pairing terms of boson operators. One must then extend the space of the Hamiltonian, such that the spinor $\Psi(\bf r)$ is written as $\Psi({\bf r}) = [\nu_{1}({\bf r}),...,\nu_{N}({\bf r}), \nu^{\dag}_{1}({\bf r}),...,\nu^{\dag}_{N}({\bf r}) ]^{\mathrm{T}}$.
  
The Hamiltonian in ${\bf k}-$space can be diagonalized with a help of a paraunitry matrix $T_{\bf k}$, such that 
\begin{align}
 T_{{\bf k}}^{\dag} {\hat H}_{\bf k} T_{{\bf k}} =  \varepsilon_{{\bf k}} =
\left[ \begin{array}{cc} E_{{\bf k}} & 0 \\ 0 & E_{-{\bf k}}  \end{array} \right],
\end{align}
where $E_{\bf k}$ is a $N\times N$ diagonal matrix of eigenvalues.
Paraunitarity of the matrix $T_{\bf k}$  means that it has to satisfy a condition  $T_{\bf k}^{\dag}\sigma_{3}T_{\bf k} = \sigma_{3}$.

We will be interested in responses of the system to external temperature gradient.
To treat the temperature gradient we adopt the Luttinger method \cite{Luttinger:PR1964} and add gravitational potentials to the Hamiltonian as
\begin{align}\label{hamtotal}
H = \frac{1}{2} \int d{\bf r}{\tilde \Psi}^{\dag}({\bf r}) {\hat H} {\tilde \Psi}({\bf r}), 
\end{align}
where ${\tilde \Psi}({\bf r}) = \left( 1+\frac{{\bf r}{\bm \nabla}\chi}{2} \right)\Psi({\bf r})$ with ${{\bm \nabla}\chi}$ being the temperature gradient with $\chi({\bf r}) =  -T({\bf r})/T$.

Let us now introduce an arbitrary operator ${\hat O}$ acting in the Hilbert space of the studied system. Density of such an operator is
${\cal O}({\bf r})= \frac{1}{2}\Psi^{\dag}({\bf r}){\hat O}\Psi({\bf r})$. Time evolution of the density is derived through a commutator with total Hamiltonian as, see Supplemental Material (SM) for details, follows
\begin{align}\label{O_continuity}
\frac{\partial {\cal O}({\bf r})}{\partial t} &= i[ H,{\cal O}({\bf r})] 
\nonumber
\\
&
= - \frac{1}{2}{\bm \nabla} {\tilde \Psi}^{\dag}({\bf r})\left( {\hat {\bf v}}\sigma_{3}{\hat O} + {\hat O}\sigma_{3}{\hat {\bf v}} \right){\tilde \Psi}({\bf r})
\nonumber
\\
&
- i\frac{1}{2}{\tilde \Psi}^{\dag}({\bf r})\left( {\hat O}\sigma_{3}{\hat H} - {\hat H}\sigma_{3}{\hat O}  \right){\tilde \Psi}({\bf r}),
\end{align}
where ${\hat {\bf v}} = i[{\hat H},{\bf r} ]$ is the velocity operator, and $\sigma_{3}$ is the third Pauli matrix operating in the extended space of the Hamiltonian (\ref{hamaf}). In deriving we assumed that the operator ${\hat O}$ commutes with the position operator. 
From (\ref{O_continuity}) we observe that for the current of an operator ${\hat O}$ to be well defined, a 
\begin{align}\label{O_condition}
{\hat O}\sigma_{3}{\hat H} - {\hat H}\sigma_{3}{\hat O} = 0
\end{align}
 condition must be satisfied by the operator ${\hat O}$. Otherwise the quantity associated  with the density ${\cal O}({\bf r})$ will not be conserved in our system. 
Let us assume we have found such an operator that satisfies the condition (\ref{O_condition}), the current associated with this operator is then defined as
\begin{align}\label{O_current}
{\bf j}_{\mathrm{O}}({\bf r}) =  {\tilde \Psi}^{\dag}({\bf r})  {\hat O}\sigma_{3}{\hat {\bf v}}  {\tilde \Psi}({\bf r}).
\end{align} 

Let us now calculate the response of the ${\hat O}-$operator current to the temperature gradient.  We will be working with the macroscopic currents, defined as ${\bf J}_{\mathrm{O}} = \frac{1}{V}\int d{\bf r} {\bf j}_{\mathrm{O}}({\bf r}) $, where $V$ is volume of the system.
Note that the current consists of unperturbed part
${\bf J}_{\mathrm{O}}^{[0]} = \frac{1}{V}\int d{\bf r}   \Psi^{\dag}({\bf r})  {\hat O}\sigma_{3}{\bf v}   \Psi({\bf r})$
and a perturbed by a temperature gradient ${\bf J}_{\mathrm{O}}^{[1]} =  \frac{1}{2V}\int d{\bf r} 
\Psi^{\dag}({\bf r}){\hat O}\sigma_{3}
\left( r_{\beta} {\hat {\bf v}} +{\hat {\bf v}} r_{\beta} \right)
\Psi({\bf r}) \nabla_{\beta} \chi$ part. 
Both of them must be used to calculate linear response to the temperature gradient. The total current is
\begin{align}
{\bf J}_{\mathrm{O}} = \left<{\bf J}_{\mathrm{O}}^{[0]}\right>_{\mathrm{ne}} + \left<{\bf J}_{\mathrm{O}}^{[1]}\right>_{\mathrm{eq}} .
\end{align}
The first term is evaluated with respect to nonequilibrium states and can be conveniently captured by the Kubo linear response formalism. Second current corresponds to orbital
magnetization in the system and it is evaluated with respect
to equilibrium state. To calculate the latter, we adopt Smrcka and Streda approach \cite{Smrcka.Streda:JPC1977} and adopt derivations presented in \cite{Matsumoto.Shindou.ea:PRB2014}.  It is important to note that the velocity written in the diagonal basis as   
$
{\tilde v}_{\alpha {\bf k}} = T_{\bf k}^{\dag}{\hat v}_{\alpha}T_{\bf k} = \partial_{\alpha} \varepsilon_{\bf k}  
+ \mathcal{A}_{\alpha {\bf k}} \sigma_{3}\varepsilon_{\bf k}  - \varepsilon_{\bf k} \sigma_{3} \mathcal{A}_{\alpha {\bf k}},
$
is conveniently separated into diagonal and non-diagonal parts,
where $\mathcal{A}_{\alpha \bf k } = T_{\bf k}^{\dag}\sigma_{3}\partial_{\alpha}T_{\bf k}$. The latter is responsible for the transverse responses of the system.
The details of the calculations for the current are given in SM.  
Overall, the total current is derived to be  
\begin{align}\label{O_current_response}
\left[{\bf J}_{\mathrm{O}} \right]_{\alpha} = \frac{1}{V}\sum_{{\bf k}n} [ {\bar \Omega}^{[\mathrm{O}]}_{\alpha\beta}({\bf k}) ]_{nn} 
 c_{1}\left[ \left(\sigma_{3}\varepsilon_{\bf k}\right)_{nn} \right]   \nabla_{\beta}\chi ,
\end{align}
where $c_{1}(x) = \int_{0}^{x} d\eta~\eta\frac{dg(\eta)}{d\eta}$, and $g(\eta) = (e^{\eta/T} - 1)^{-1}$ is the Bose-Einstein distribution function. We defined an ${\mathrm O}-$Berry curvature,
\begin{align}\label{O_Berry}
{\bar \Omega}^{[\mathrm{O}]}_{\alpha\beta}({\bf k}) = i {\bar O}\partial_{\alpha}T_{\bf k}^{\dag}\sigma_{3}\partial_{\beta}T_{\bf k}  
- \left( \alpha \leftrightarrow \beta\right),
\end{align}
a Berry curvature modified with an operator ${\bar O} = \sigma_{3}T_{\bf k}^{\dag}{\hat O}T_{\bf k}\sigma_{3}$. Due to commutation relations (\ref{O_condition}), matrix ${\bar O}$ is diagonal in band index. We show there is a sum rule $\sum_{n} [ {\bar \Omega}^{[\mathrm{O}]}_{\alpha\beta}({\bf k}) ]_{nn} = 0$ the ${\mathrm O}-$Berry curvature satisfies. Expressions (\ref{O_current_response}) and (\ref{O_Berry}) together with (\ref{O_condition}) and (\ref{O_current}) are the main results of this paper.

\noindent

\textit{\underline{Single layer honeycomb antiferromagnet}.} We now apply our results to specific model of an antiferromagnet on honeycomb lattice. 
The lattice of the system is shown in Fig.~\ref{fig:honeycomb}. We define an exchange Hamiltonian
\begin{align}
H = J\sum_{<ij>}{\bf S}_{i}{\bf S}_{j} +  D \sum_{<<ij>>} \nu_{ij}\left[ {\bf S}_{i}\times {\bf S}_{j}\right]_{z}.
\end{align} 
Here $J>0$ is the nearest neighbor spin exchange, $D$ is the strength of the second-nearest neighbor spin DMI, and $\nu_{ij}$ is a sign convention defined in Fig.~\ref{fig:honeycomb}.  

Let us assume there is a Neel order in the direction perpendicular to lattice plane, $z-$direction. 
To study magnons of the model we perform Holstein-Primakoff transformation from spins to boson operators, $S_{A+} =\sqrt{2S-a^{\dag}a} a$, $S_{Az} = S - a^{\dag}a$, and $S_{B+}= -\sqrt{2S - b^{\dag}b}b^{\dag}$, $S_{Bz}= -S + b^{\dag}b$, and assume large $S$ limit. As shown in SM, the Hamiltonian describing non-interacting magnons splits in to two blocks. The first block, call it $\mathrm{\Rmnum{1}}$, is described by $\Psi_{\mathrm{\Rmnum{1}}} = ( a_{\bf k}, b^{\dag}_{-\bf k} )^{\mathrm{T}}$ spinor. The Fourier image of the Hamiltonian of the first block is
\begin{align}\label{hamfirst}
H_{\mathrm{\Rmnum{1}} \bf k} =  JS \left[
\begin{array}{cc}
3 +\Delta_{\bf k}  & - \gamma_{\bf k} \\
- \gamma_{-\bf k} & 3 - \Delta_{\bf k}
\end{array}
\right].
\end{align}
where we defined $\gamma_{\bf k} = 2e^{i \frac{k_{x}}{2\sqrt{3}}}\cos( \frac{k_{y}}{2} ) + e^{-i\frac{k_{x}}{\sqrt{3}}}$, and $\Delta_{{\bf k}} = 2\frac{D}{J} [\sin(k_{y}) - 2\sin( \frac{k_{y}}{2})\cos(\frac{\sqrt{3}k_{x}}{2} ) ]$ is the DMI, and we note $\Delta_{{\bf k}} = - \Delta_{-{\bf k}}$. Hamiltonian of the second block described by $\Psi_{\mathrm{\Rmnum{2}}} = ( b_{\bf k}, a^{\dag}_{-\bf k} )^{\mathrm{T}}$ spinor is obtained by $\gamma_{\bf k} \rightarrow \gamma_{-\bf k}$ in (\ref{hamfirst}).

Let us define operator ${\hat O}$  acting in full, $\Psi_{\bf k} =  (a_{\bf k}, b_{\bf k}, a^{\dag}_{-\bf k},b^{\dag}_{-\bf k} )^{\mathrm{T}}$, space as 
\begin{align}
{\hat O} = \left[
\begin{array}{cc}
{\hat \tau}_{3} & 0 \\
0 & {\hat \tau}_{3} 
\end{array} \right],
\end{align} 
where ${\hat \tau}_{3}$ is third $2\times 2$ Pauli matrix. The density of this operator written in real space, ${\cal O}({\bf r}) = \frac{1}{2}\Psi^{\dag}({\bf r}){\hat O}\Psi({\bf r}) = a^{\dag}({\bf r})a({\bf r})- b^{\dag}({\bf r})b({\bf r}) $, is the spin density. It can be shown that such an operator satisfies condition (\ref{O_condition}), thus the spin density current associated with ${\hat O}$ is well defined. Let us now calculate the spin density current as a response to the temperature gradient. Expression for the response is given by (\ref{O_current_response}), hence we need to find eigenvalues and calculate $\mathrm{O}-$Berry curvature.

Spectrum of magnons for both blocks of the Hamiltonian is obtained to be
\begin{align}
E_{ \bf k} =JS\left(  \Delta_{\bf k}  +  \sqrt{ 9 - \vert \gamma_{\bf k} \vert^2  } \right).
\end{align}
Paraunitary matrix $T_{ \mathrm{\Rmnum{1}} {\bf k}}$ that diagonalizes the Hamiltonian is readily constructed to be
\begin{align}
T_{ \mathrm{\Rmnum{1}} {\bf k}} = \left[
\begin{array}{cc}
\cosh(\xi_{  {\bf k} }/2 ) e^{i\chi_{{\bf k}}} & \sinh(\xi_{  {\bf k}}/2 )   \\
\sinh(\xi_{ {\bf k}}/2 )  & \cosh(\xi_{  {\bf k} }/2 ) e^{-i\chi_{{\bf k}}}
\end{array} \right],
\end{align}
where $\sinh(\xi_{  {\bf k}}) = \vert\gamma_{\bf k}\vert/\sqrt{9 - \vert \gamma_{\bf k}\vert^2}$, $\cosh(\xi_{  {\bf k}}) = 3/\sqrt{9 - \vert \gamma_{\bf k}\vert^2}$, and $\gamma_{\bf k} = \vert \gamma_{\bf k} \vert e^{i\chi_{\bf k}}$. One can show that the $\mathrm{\Rmnum{2}}$ block described by $\Psi_{\mathrm{\Rmnum{2}}} = ( b_{\bf k}, a^{\dag}_{-\bf k} )^{\mathrm{T}}$ spinor has the paraunitary matrix $T_{ \mathrm{\Rmnum{2}} {\bf k}}$ obtained from the $T_{ \mathrm{\Rmnum{1}} {\bf k}}$ by setting $\chi_{\bf k} \rightarrow - \chi_{\bf k}$, and hence has the same ${\mathrm O}-$Berry curvature (see SM for more details). 
The spin density current can then be written as 
\begin{align}\label{current_single}
\left[ {\bf J}_{\mathrm{O}} \right]_{\alpha} =- \frac{1}{V}\sum_{\bf k} 2 \Omega^{[\mathrm{O}]}_{\alpha\beta}({\bf k})\left[ c_{1}(E_{\bf k}) - c_{1}(E_{-\bf k})\right] \nabla_{\beta}\chi ,
\end{align}
with the diagonal elements of the $\mathrm{O}-$Berry curvature written as
\begin{align}
 \Omega_{\alpha\beta}^{[\mathrm{O}]}({\bf k})
&
=   -\frac{3}{2\left( 9 - \vert \gamma_{\bf k}\vert^2 \right)^{3/2}}
\\
&
\times\left[ \left( \partial_{\alpha}\mathrm{Re}\gamma_{\bf k}\right)\left( \partial_{\beta}\mathrm{Im}\gamma_{\bf k}\right) 
- \left(\partial_{\beta}\mathrm{Re}\gamma_{\bf k} \right)\left(\partial_{\alpha}\mathrm{Im}\gamma_{\bf k} \right)   \right].
\nonumber
\end{align}
We observe that the current vanishes if the DMI is zero in the system, in which case $E_{\bf k} = E_{-\bf k}$. Note that the $\mathrm{O}-$Berry curvature is independent of the DMI.

\begin{figure}[t] \includegraphics[width=1\linewidth]{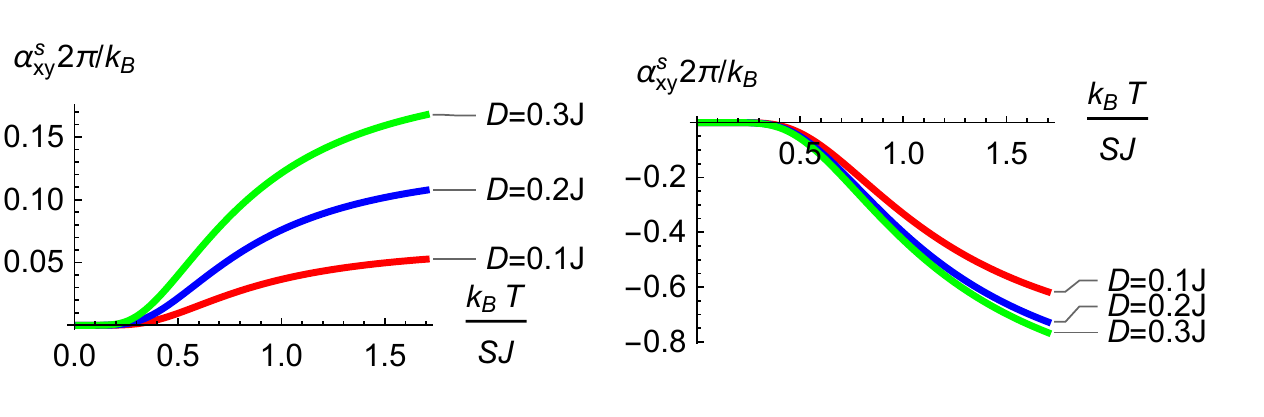}

\protect\caption{(Color online) Spin Nernst conductivity $\alpha^{\mathrm{s}}_{xy}$, defined after expression (\ref{current_single}). Left: a single layer honeycomb antiferromagnet. Right: double layer honeycomb antiferromagnet. Plots are given for different values of DMI.}

\label{Current}  

\end{figure}
Recalling the definition of $\chi(\bf r)$, we define SNE conductivity $\alpha^{\mathrm{s}}_{\alpha\beta}$ as $\left[{\bf J}_{\mathrm{O}} \right]_{\alpha} = -\alpha^{\mathrm{s}}_{\alpha\beta}\nabla_{\beta}T(\bf r)$, and plot its dependence on the temperature - see Fig. \ref{Current}. 
We now wish to extract analytic results in the limit of small DMI, $D<J$. 
There are two different symmetry points, namely ${\bm \Gamma}$, and ${\bf K}$, ${\bf K}^\prime$ points, in the Brillouin zone of magnons the spin current gets major contributions from.
Close to the ${\bm \Gamma} = (0,0)$ point the spectrum is ungapped and linear.
We expand all functions close to the ${\bm \Gamma}$ point to obtain a low temperature, $T<JS$, dependence of the current. See SM for details. 
\begin{align}
\left[ \left( {\bf J}_{\mathrm{O}} \right)_{x} \right]_{{\bm \Gamma}} = \frac{5 \zeta(5) }{9\sqrt{3}\pi V} \frac{D}{J} \left( \frac{T}{JS} \right)^{4} \nabla_{y}T(\bf r) ,
\end{align} 
where an estimate of Riemann zeta function is $\zeta(5) \approx 1$. At ${\bf K} =(0,- 4\pi/3) $ and ${\bf K}^\prime = (0,+ 4\pi/3)$ points, the Berry curvature has an absolute value maximum. 
An analytic estimate of the current contribution from these points at small temperatures $T<JS$, is obtained
$
\left[ \left( {\bf J}_{\mathrm{O}} \right)_{x} \right]_{{\bf K}} =  \frac{9\sqrt{3}\Lambda^2}{8\pi V}\frac{D}{J} \left(\frac{JS}{T}\right)^2 e^{-\frac{3JS}{T}} \nabla_{y}T(\bf r) ,
$
where we introduced a high limit cut-off $\Lambda \sim 1$ for $k$, such that $\sum_{{\bf k}} = \frac{\Lambda^2}{4\pi}$. It is straightforward to show that $\left[ \left( J_{\mathrm{O}} \right)_{x} \right]_{{\bm \Gamma}} \gg \left[ \left( J_{\mathrm{O}} \right)_{x} \right]_{{\bf K}}$ for small temperatures. Both contributions are of the same sign which always results in the same sign of SNE for this model irrespective of the temperature and the strength of DMI.

\begin{figure} \centerline{\includegraphics[clip,width=1\columnwidth]{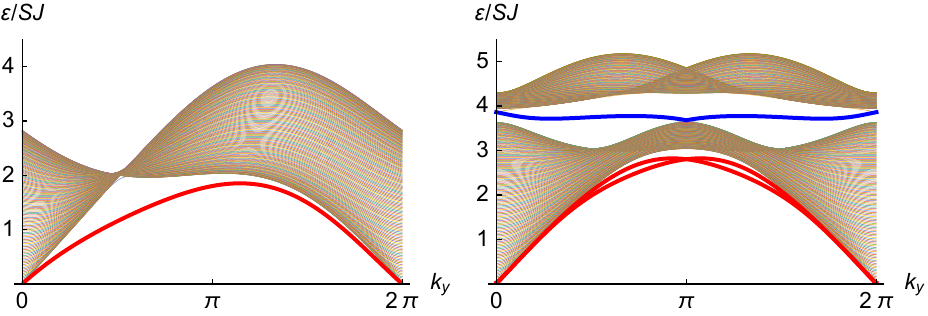}}

\protect\caption{(Color online) 
Magnon spectrum of 80 atoms wide strip of honeycomb lattice antiferromagnet. 
Strip is in $x-$direction, while $y-$direction is assumed infinite.
The edges of the system are of the zig-zag type.
Left: Single layer with DMI, $D=0.2J$.
Right: Double layer.  Protected magnon edge states occur in high energy band gap.
Parameters are chosen to be $J^{\prime} = 1.3J$ and $D=0.2J$. }
\label{fig:Strip}  

\end{figure}

The Chern number of the magnon band for the single layer honeycomb antiferromagnet is zero (see Fig.~\ref{fig:honeycomb}). As a result we do not observe any protected by the Chern number edge states in the finite strip geometry with a zig-zag edge (see Fig.~\ref{fig:Strip}). Nevertheless, we observe an edge state  analogous to the zero energy edge state in fermionic model of graphene with a zig-zag or bearded edge. The edge state connects ${\bf K}$ and ${\bf K}^\prime$ points which have different in sign Berry curvatures.  Such edge
states do not contribute to the SNE in the finite geometry
of a single layer honeycomb antiferromagnet.

\noindent
\textit{\underline{Double layer honeycomb antiferromagnet}.} 
In another model we consider an antiferromagnet on a double layer honeycomb lattice (see Fig.~\ref{fig:honeycomb}). We again assume nearest neighbor antiferromagnetic exchange interaction,  second-nearest neighbor DMI, same in both layers, and antiferromagnetic interaction between the layers denoted by $J^\prime$. 
With the Neel order being in $z-$direction, we follow the same steps, as in the previous example, 
and get spectrum of spin waves 
\begin{align}
 E_{\bf k \pm }^{2}/(SJ)^2  
&= \lambda^2 - \vert \gamma_{\bf k} \vert^2 + \Delta_{\bf k}^{2} -t^2 \nonumber
\\
&
\pm 2 \sqrt{\Delta_{\bf k}^{2}(\lambda^2 - \vert\gamma_{\bf k}\vert^2) + t^2\vert\gamma_{\bf k}\vert^2},
\end{align}
here $\lambda = 3+t$, where $t = J^{\prime}/J $.
The spectrum and the Berry curvature distribution is shown in Fig. \ref{fig:honeycomb}.
There we observe that the Berry curvature is of the monopole type located at the ${\bf M}$ points in the Brillouin zone in contrast to the magnon Haldane-Kane-Mele model \cite{Kim.Ochoa.ea:apa2016}.

For this model the Chern numbers of the upper and lower bands are $+1$ and $-1$, respectively, where the topological charge is $1/3$ per $\bf M$ point. The whole band now contributes in an additive way to SNE which results in a much larger effect. Numerical calculations of the magnon SNE are shown in Fig.~\ref{Current}.  
To uncover the role of the edge states, we calculate the energy spectrum of a double-layer strip with a zig-zag edge, see Fig. \ref{fig:Strip}. The high-energy edge states here are due to DMI, in contrast to the single-layer model. These edge states are chiral and are protected by the finite Chern number due to the non-trivial topology of the bulk magnons. These edge states are also expected to contribute to SNE conductivity in the finite geometry \cite{Matsumoto.Murakami:PRL2011}. The low-energy edge states are of the same nature as in single layer honeycomb antiferromagnet and are not expected to contribute to SNE.

\textit{\underline{Absence of thermal Hall effect}.} The thermal Hall coefficient is given by an expression $\kappa_{xy} = -\frac{1}{2T}\sum_{\bf k}\sum_{n=1}^{2N}\left[ \Omega_{xy}(\bf k)\right]_{nn}c_{2}\left[ (\sigma_{3}\varepsilon_{\bf k})_{nn}\right]$, where we defined $c_{2}(x) = \int_{0}^{x}d\eta ~\eta^2\frac{dg}{d\eta}$.
We set ${\hat O} = \sigma_{3}$ in expression (\ref{O_Berry}), to obtain the Berry curvature of the energy bands $
\Omega_{xy}({\bf k}) = i \sigma_{3}\partial_{x}T_{\bf k}^{\dag}\sigma_{3}\partial_{y}T_{\bf k}  
- \left( x \leftrightarrow y \right)$. 
For an antiferromagnet on a single layer honeycomb lattice, the energy states are degenerate, corresponding to the two blocks, $\mathrm{\Rmnum{1}}$ and $\mathrm{\Rmnum{2}}$, with opposite in sign Berry curvatures. The two blocks correspond to two sublattices related either by inversion $\mathcal{I}$ or by time-reversal $\mathcal{T}$ transformations.  On the other hand, the double layer antiferromagnet in Fig.~\ref{fig:honeycomb} is invariant under the global time reversal symmetry if treated as a 2D system since $\mathcal{T}$ followed by interchange of honeycomb layers is a symmetry. Thus, the thermal Hall response considered in \cite{Matsumoto.Shindou.ea:PRB2014} vanishes for both models in Fig.~\ref{fig:honeycomb}. 

\textit{\underline{Conclusions}.} In this paper we theoretically studied magnon mediated SNE in antiferromagnets. We gave a general condition for a current to be a well-defined quantity in an antiferromagnet, and then derived its response to external temperature gradient. We showed that transverse response of this current is defined by a modified Berry curvature. In antiferromagnets with Neel order, SNE can be driven by the Dzyaloshinskii-Moriya interaction and SNE is present even in systems with $\mathcal{T}\mathcal{I}$ or global $\mathcal{T}$ symmetries. In both cases the thermal Hall effect is zero while SNE should change sign with the reversal of the Neel vector in the former case but not in the latter case. We also identified the protected edge states with counterpropagating magnon modes, carrying spin but no energy.

\textit{Acknowledgements.} We gratefully acknowledge useful
discussions with K.~Belashchenko. This work was supported primarily by the DOE Early Career Award DE-SC0014189.

\textit{Note added.} During the completion of the work, see \cite{Zyuzin.Kovalev.MM2016}, we became aware of a Letter \cite{Cheng.Okamoto.Xiao:ArXiv} that discusses SNE in antiferromagnets. 
We believe the two Letters compliment each other.

\bibliographystyle{apsrev}
\bibliography{MyBIB}

\begin{thebibliography}{50}
\expandafter\ifx\csname natexlab\endcsname\relax\def\natexlab#1{#1}\fi
\expandafter\ifx\csname bibnamefont\endcsname\relax
  \def\bibnamefont#1{#1}\fi
\expandafter\ifx\csname bibfnamefont\endcsname\relax
  \def\bibfnamefont#1{#1}\fi
\expandafter\ifx\csname citenamefont\endcsname\relax
  \def\citenamefont#1{#1}\fi
\expandafter\ifx\csname url\endcsname\relax
  \def\url#1{\texttt{#1}}\fi
\expandafter\ifx\csname urlprefix\endcsname\relax\def\urlprefix{URL }\fi
\providecommand{\bibinfo}[2]{#2}
\providecommand{\eprint}[2][]{\url{#2}}

\bibitem[{\citenamefont{Dyakonov}(2008)}]{Dyakonov}
\bibinfo{editor}{\bibfnamefont{M.~I.} \bibnamefont{Dyakonov}}, ed.,
  \emph{\bibinfo{title}{Spin Physics in Semiconductors}}
  (\bibinfo{publisher}{Springer-Verlag Berlin Heidelberg},
  \bibinfo{year}{2008}).

\bibitem[{\citenamefont{{{\v Z}uti{\'c}} et~al.}(2004)\citenamefont{{{\v
  Z}uti{\'c}}, {Fabian}, and {Das Sarma}}}]{Zutic:RoMP2004}
\bibinfo{author}{\bibfnamefont{I.}~\bibnamefont{{{\v Z}uti{\'c}}}},
  \bibinfo{author}{\bibfnamefont{J.}~\bibnamefont{{Fabian}}}, \bibnamefont{and}
  \bibinfo{author}{\bibfnamefont{S.}~\bibnamefont{{Das Sarma}}},
  \bibinfo{journal}{Rev. Mod. Phys.} \textbf{\bibinfo{volume}{76}},
  \bibinfo{pages}{323} (\bibinfo{year}{2004}).

\bibitem[{\citenamefont{{Bader} and {Parkin}}(2010)}]{Bader.Parkin:ARoCMP2010}
\bibinfo{author}{\bibfnamefont{S.~D.} \bibnamefont{{Bader}}} \bibnamefont{and}
  \bibinfo{author}{\bibfnamefont{S.~S.~P.} \bibnamefont{{Parkin}}},
  \bibinfo{journal}{Annual Review of Condensed Matter Physics}
  \textbf{\bibinfo{volume}{1}}, \bibinfo{pages}{71} (\bibinfo{year}{2010}).

\bibitem[{\citenamefont{{Dyakonov} and {Perel}}(1971)}]{Dyakonov.Perel:PLA1971}
\bibinfo{author}{\bibfnamefont{M.~I.} \bibnamefont{{Dyakonov}}}
  \bibnamefont{and} \bibinfo{author}{\bibfnamefont{V.~I.}
  \bibnamefont{{Perel}}}, \bibinfo{journal}{Phys. Lett. A}
  \textbf{\bibinfo{volume}{35}}, \bibinfo{pages}{459} (\bibinfo{year}{1971}).

\bibitem[{\citenamefont{{Hirsch}}(1999)}]{Hirsch:PRL1999}
\bibinfo{author}{\bibfnamefont{J.~E.} \bibnamefont{{Hirsch}}},
  \bibinfo{journal}{Phys. Rev. Lett.} \textbf{\bibinfo{volume}{83}},
  \bibinfo{pages}{1834} (\bibinfo{year}{1999}).

\bibitem[{\citenamefont{{Zhang}}(2000)}]{Zhang:PRL2000}
\bibinfo{author}{\bibfnamefont{S.}~\bibnamefont{{Zhang}}},
  \bibinfo{journal}{Phys. Rev. Lett.} \textbf{\bibinfo{volume}{85}},
  \bibinfo{pages}{393} (\bibinfo{year}{2000}).

\bibitem[{\citenamefont{{Murakami} et~al.}(2003)\citenamefont{{Murakami},
  {Nagaosa}, and {Zhang}}}]{Murakami:S2003}
\bibinfo{author}{\bibfnamefont{S.}~\bibnamefont{{Murakami}}},
  \bibinfo{author}{\bibfnamefont{N.}~\bibnamefont{{Nagaosa}}},
  \bibnamefont{and} \bibinfo{author}{\bibfnamefont{S.-C.}
  \bibnamefont{{Zhang}}}, \bibinfo{journal}{Science}
  \textbf{\bibinfo{volume}{301}}, \bibinfo{pages}{1348} (\bibinfo{year}{2003}).

\bibitem[{\citenamefont{{Sinova} et~al.}(2004)\citenamefont{{Sinova}, {Culcer},
  {Niu}, {Sinitsyn}, {Jungwirth}, and {MacDonald}}}]{Sinova:PRL2004}
\bibinfo{author}{\bibfnamefont{J.}~\bibnamefont{{Sinova}}},
  \bibinfo{author}{\bibfnamefont{D.}~\bibnamefont{{Culcer}}},
  \bibinfo{author}{\bibfnamefont{Q.}~\bibnamefont{{Niu}}},
  \bibinfo{author}{\bibfnamefont{N.~A.} \bibnamefont{{Sinitsyn}}},
  \bibinfo{author}{\bibfnamefont{T.}~\bibnamefont{{Jungwirth}}},
  \bibnamefont{and} \bibinfo{author}{\bibfnamefont{A.~H.}
  \bibnamefont{{MacDonald}}}, \bibinfo{journal}{Phys. Rev. Lett.}
  \textbf{\bibinfo{volume}{92}}, \bibinfo{eid}{126603} (\bibinfo{year}{2004}).

\bibitem[{\citenamefont{{Kato} et~al.}(2004)\citenamefont{{Kato}, {Myers},
  {Gossard}, and {Awschalom}}}]{Kato.Myers.ea:S2004}
\bibinfo{author}{\bibfnamefont{Y.~K.} \bibnamefont{{Kato}}},
  \bibinfo{author}{\bibfnamefont{R.~C.} \bibnamefont{{Myers}}},
  \bibinfo{author}{\bibfnamefont{A.~C.} \bibnamefont{{Gossard}}},
  \bibnamefont{and} \bibinfo{author}{\bibfnamefont{D.~D.}
  \bibnamefont{{Awschalom}}}, \bibinfo{journal}{Science}
  \textbf{\bibinfo{volume}{306}}, \bibinfo{pages}{1910} (\bibinfo{year}{2004}).

\bibitem[{\citenamefont{{Valenzuela} and
  {Tinkham}}(2006)}]{Valenzuela.Tinkham:N2006}
\bibinfo{author}{\bibfnamefont{S.~O.} \bibnamefont{{Valenzuela}}}
  \bibnamefont{and}
  \bibinfo{author}{\bibfnamefont{M.}~\bibnamefont{{Tinkham}}},
  \bibinfo{journal}{Nature} \textbf{\bibinfo{volume}{442}},
  \bibinfo{pages}{176} (\bibinfo{year}{2006}).

\bibitem[{\citenamefont{{Sinova} et~al.}(2015)\citenamefont{{Sinova},
  {Valenzuela}, {Wunderlich}, {Back}, and
  {Jungwirth}}}]{Sinova.Valenzuela.ea:RoMP2015}
\bibinfo{author}{\bibfnamefont{J.}~\bibnamefont{{Sinova}}},
  \bibinfo{author}{\bibfnamefont{S.~O.} \bibnamefont{{Valenzuela}}},
  \bibinfo{author}{\bibfnamefont{J.}~\bibnamefont{{Wunderlich}}},
  \bibinfo{author}{\bibfnamefont{C.~H.} \bibnamefont{{Back}}},
  \bibnamefont{and}
  \bibinfo{author}{\bibfnamefont{T.}~\bibnamefont{{Jungwirth}}},
  \bibinfo{journal}{Rev. Mod. Phys.} \textbf{\bibinfo{volume}{87}},
  \bibinfo{pages}{1213} (\bibinfo{year}{2015}).

\bibitem[{\citenamefont{Kane and Mele}(2005)}]{Kane.Mele:PRL2005}
\bibinfo{author}{\bibfnamefont{C.~L.} \bibnamefont{Kane}} \bibnamefont{and}
  \bibinfo{author}{\bibfnamefont{E.~J.} \bibnamefont{Mele}},
  \bibinfo{journal}{Phys. Rev. Lett.} \textbf{\bibinfo{volume}{95}},
  \bibinfo{pages}{146802} (\bibinfo{year}{2005}).

\bibitem[{\citenamefont{{Bernevig} et~al.}(2006)\citenamefont{{Bernevig},
  {Hughes}, and {Zhang}}}]{Bernevig.Hughes.ea:S2006}
\bibinfo{author}{\bibfnamefont{B.~A.} \bibnamefont{{Bernevig}}},
  \bibinfo{author}{\bibfnamefont{T.~L.} \bibnamefont{{Hughes}}},
  \bibnamefont{and} \bibinfo{author}{\bibfnamefont{S.-C.}
  \bibnamefont{{Zhang}}}, \bibinfo{journal}{Science}
  \textbf{\bibinfo{volume}{314}}, \bibinfo{pages}{1757} (\bibinfo{year}{2006}).

\bibitem[{\citenamefont{{Miron} et~al.}(2011)\citenamefont{{Miron}, {Garello},
  {Gaudin}, {Zermatten}, {Costache}, {Auffret}, {Bandiera}, {Rodmacq},
  {Schuhl}, and {Gambardella}}}]{Miron.Garello.ea:N2011}
\bibinfo{author}{\bibfnamefont{I.~M.} \bibnamefont{{Miron}}},
  \bibinfo{author}{\bibfnamefont{K.}~\bibnamefont{{Garello}}},
  \bibinfo{author}{\bibfnamefont{G.}~\bibnamefont{{Gaudin}}},
  \bibinfo{author}{\bibfnamefont{P.-J.} \bibnamefont{{Zermatten}}},
  \bibinfo{author}{\bibfnamefont{M.~V.} \bibnamefont{{Costache}}},
  \bibinfo{author}{\bibfnamefont{S.}~\bibnamefont{{Auffret}}},
  \bibinfo{author}{\bibfnamefont{S.}~\bibnamefont{{Bandiera}}},
  \bibinfo{author}{\bibfnamefont{B.}~\bibnamefont{{Rodmacq}}},
  \bibinfo{author}{\bibfnamefont{A.}~\bibnamefont{{Schuhl}}}, \bibnamefont{and}
  \bibinfo{author}{\bibfnamefont{P.}~\bibnamefont{{Gambardella}}},
  \bibinfo{journal}{Nature} \textbf{\bibinfo{volume}{476}},
  \bibinfo{pages}{189} (\bibinfo{year}{2011}).

\bibitem[{\citenamefont{{Liu} et~al.}(2012)\citenamefont{{Liu}, {Lee},
  {Gudmundsen}, {Ralph}, and {Buhrman}}}]{Liu.Lee.ea:PRL2012}
\bibinfo{author}{\bibfnamefont{L.}~\bibnamefont{{Liu}}},
  \bibinfo{author}{\bibfnamefont{O.~J.} \bibnamefont{{Lee}}},
  \bibinfo{author}{\bibfnamefont{T.~J.} \bibnamefont{{Gudmundsen}}},
  \bibinfo{author}{\bibfnamefont{D.~C.} \bibnamefont{{Ralph}}},
  \bibnamefont{and} \bibinfo{author}{\bibfnamefont{R.~A.}
  \bibnamefont{{Buhrman}}}, \bibinfo{journal}{Phys. Rev. Lett.}
  \textbf{\bibinfo{volume}{109}}, \bibinfo{eid}{096602} (\bibinfo{year}{2012}).

\bibitem[{\citenamefont{Liu et~al.}(2012)\citenamefont{Liu, Pai, Li, Tseng,
  Ralph, and Buhrman}}]{Liu:Science2012}
\bibinfo{author}{\bibfnamefont{L.}~\bibnamefont{Liu}},
  \bibinfo{author}{\bibfnamefont{C.-F.} \bibnamefont{Pai}},
  \bibinfo{author}{\bibfnamefont{Y.}~\bibnamefont{Li}},
  \bibinfo{author}{\bibfnamefont{H.~W.} \bibnamefont{Tseng}},
  \bibinfo{author}{\bibfnamefont{D.~C.} \bibnamefont{Ralph}}, \bibnamefont{and}
  \bibinfo{author}{\bibfnamefont{R.~A.} \bibnamefont{Buhrman}},
  \bibinfo{journal}{Science} \textbf{\bibinfo{volume}{336}},
  \bibinfo{pages}{555} (\bibinfo{year}{2012}).

\bibitem[{\citenamefont{{Uchida} et~al.}(2008)\citenamefont{{Uchida},
  {Takahashi}, {Harii}, {Ieda}, {Koshibae}, {Ando}, {Maekawa}, and
  {Saitoh}}}]{Uchida:Nature2008}
\bibinfo{author}{\bibfnamefont{K.-I.} \bibnamefont{{Uchida}}},
  \bibinfo{author}{\bibfnamefont{S.}~\bibnamefont{{Takahashi}}},
  \bibinfo{author}{\bibfnamefont{K.}~\bibnamefont{{Harii}}},
  \bibinfo{author}{\bibfnamefont{J.}~\bibnamefont{{Ieda}}},
  \bibinfo{author}{\bibfnamefont{W.}~\bibnamefont{{Koshibae}}},
  \bibinfo{author}{\bibfnamefont{K.}~\bibnamefont{{Ando}}},
  \bibinfo{author}{\bibfnamefont{S.}~\bibnamefont{{Maekawa}}},
  \bibnamefont{and} \bibinfo{author}{\bibfnamefont{E.}~\bibnamefont{{Saitoh}}},
  \bibinfo{journal}{Nature} \textbf{\bibinfo{volume}{455}},
  \bibinfo{pages}{778} (\bibinfo{year}{2008}).

\bibitem[{\citenamefont{{Uchida} et~al.}(2010)\citenamefont{{Uchida}, {Xiao},
  {Adachi}, {Ohe}, {Takahashi}, {Ieda}, {Ota}, {Kajiwara}, {Umezawa}, {Kawai}
  et~al.}}]{Uchida.Xiao.ea:NM2010}
\bibinfo{author}{\bibfnamefont{K.}~\bibnamefont{{Uchida}}},
  \bibinfo{author}{\bibfnamefont{J.}~\bibnamefont{{Xiao}}},
  \bibinfo{author}{\bibfnamefont{H.}~\bibnamefont{{Adachi}}},
  \bibinfo{author}{\bibfnamefont{J.}~\bibnamefont{{Ohe}}},
  \bibinfo{author}{\bibfnamefont{S.}~\bibnamefont{{Takahashi}}},
  \bibinfo{author}{\bibfnamefont{J.}~\bibnamefont{{Ieda}}},
  \bibinfo{author}{\bibfnamefont{T.}~\bibnamefont{{Ota}}},
  \bibinfo{author}{\bibfnamefont{Y.}~\bibnamefont{{Kajiwara}}},
  \bibinfo{author}{\bibfnamefont{H.}~\bibnamefont{{Umezawa}}},
  \bibinfo{author}{\bibfnamefont{H.}~\bibnamefont{{Kawai}}},
  \bibnamefont{et~al.}, \bibinfo{journal}{Nat. Mater.}
  \textbf{\bibinfo{volume}{9}}, \bibinfo{pages}{894} (\bibinfo{year}{2010}).

\bibitem[{\citenamefont{{Jaworski} et~al.}(2010)\citenamefont{{Jaworski},
  {Yang}, {Mack}, {Awschalom}, {Heremans}, and
  {Myers}}}]{Jaworski.Yang.ea:NM2010}
\bibinfo{author}{\bibfnamefont{C.~M.} \bibnamefont{{Jaworski}}},
  \bibinfo{author}{\bibfnamefont{J.}~\bibnamefont{{Yang}}},
  \bibinfo{author}{\bibfnamefont{S.}~\bibnamefont{{Mack}}},
  \bibinfo{author}{\bibfnamefont{D.~D.} \bibnamefont{{Awschalom}}},
  \bibinfo{author}{\bibfnamefont{J.~P.} \bibnamefont{{Heremans}}},
  \bibnamefont{and} \bibinfo{author}{\bibfnamefont{R.~C.}
  \bibnamefont{{Myers}}}, \bibinfo{journal}{Nat. Mater.}
  \textbf{\bibinfo{volume}{9}}, \bibinfo{pages}{898} (\bibinfo{year}{2010}).

\bibitem[{\citenamefont{{Qi} et~al.}(2006)\citenamefont{{Qi}, {Wu}, and
  {Zhang}}}]{Qi.Wu.ea:PRB2006}
\bibinfo{author}{\bibfnamefont{X.-L.} \bibnamefont{{Qi}}},
  \bibinfo{author}{\bibfnamefont{Y.-S.} \bibnamefont{{Wu}}}, \bibnamefont{and}
  \bibinfo{author}{\bibfnamefont{S.-C.} \bibnamefont{{Zhang}}},
  \bibinfo{journal}{Phys. Rev. B} \textbf{\bibinfo{volume}{74}},
  \bibinfo{eid}{085308} (\bibinfo{year}{2006}).

\bibitem[{\citenamefont{Onose et~al.}(2010)\citenamefont{Onose, Ideue, Katsura,
  Shiomi, Nagaosa, and Tokura}}]{Onose.Ideue.ea:S2010}
\bibinfo{author}{\bibfnamefont{Y.}~\bibnamefont{Onose}},
  \bibinfo{author}{\bibfnamefont{T.}~\bibnamefont{Ideue}},
  \bibinfo{author}{\bibfnamefont{H.}~\bibnamefont{Katsura}},
  \bibinfo{author}{\bibfnamefont{Y.}~\bibnamefont{Shiomi}},
  \bibinfo{author}{\bibfnamefont{N.}~\bibnamefont{Nagaosa}}, \bibnamefont{and}
  \bibinfo{author}{\bibfnamefont{Y.}~\bibnamefont{Tokura}},
  \bibinfo{journal}{Science} \textbf{\bibinfo{volume}{329}},
  \bibinfo{pages}{297} (\bibinfo{year}{2010}).

\bibitem[{\citenamefont{{Ideue} et~al.}(2012)\citenamefont{{Ideue}, {Onose},
  {Katsura}, {Shiomi}, {Ishiwata}, {Nagaosa}, and
  {Tokura}}}]{Ideue.Onose.ea:PRB2012}
\bibinfo{author}{\bibfnamefont{T.}~\bibnamefont{{Ideue}}},
  \bibinfo{author}{\bibfnamefont{Y.}~\bibnamefont{{Onose}}},
  \bibinfo{author}{\bibfnamefont{H.}~\bibnamefont{{Katsura}}},
  \bibinfo{author}{\bibfnamefont{Y.}~\bibnamefont{{Shiomi}}},
  \bibinfo{author}{\bibfnamefont{S.}~\bibnamefont{{Ishiwata}}},
  \bibinfo{author}{\bibfnamefont{N.}~\bibnamefont{{Nagaosa}}},
  \bibnamefont{and} \bibinfo{author}{\bibfnamefont{Y.}~\bibnamefont{{Tokura}}},
  \bibinfo{journal}{Phys. Rev. B} \textbf{\bibinfo{volume}{85}},
  \bibinfo{eid}{134411} (\bibinfo{year}{2012}).

\bibitem[{\citenamefont{{Katsura} et~al.}(2010)\citenamefont{{Katsura},
  {Nagaosa}, and {Lee}}}]{Katsura.Nagaosa.ea:PRL2010}
\bibinfo{author}{\bibfnamefont{H.}~\bibnamefont{{Katsura}}},
  \bibinfo{author}{\bibfnamefont{N.}~\bibnamefont{{Nagaosa}}},
  \bibnamefont{and} \bibinfo{author}{\bibfnamefont{P.~A.} \bibnamefont{{Lee}}},
  \bibinfo{journal}{Phys. Rev. Lett.} \textbf{\bibinfo{volume}{104}},
  \bibinfo{eid}{066403} (\bibinfo{year}{2010}).

\bibitem[{\citenamefont{Matsumoto and
  Murakami}(2011)}]{Matsumoto.Murakami:PRL2011}
\bibinfo{author}{\bibfnamefont{R.}~\bibnamefont{Matsumoto}} \bibnamefont{and}
  \bibinfo{author}{\bibfnamefont{S.}~\bibnamefont{Murakami}},
  \bibinfo{journal}{Phys. Rev. Lett.} \textbf{\bibinfo{volume}{106}},
  \bibinfo{pages}{197202} (\bibinfo{year}{2011}).

\bibitem[{\citenamefont{{Zhang} et~al.}(2013)\citenamefont{{Zhang}, {Ren},
  {Wang}, and {Li}}}]{Zhang.Ren.ea:PRB2013}
\bibinfo{author}{\bibfnamefont{L.}~\bibnamefont{{Zhang}}},
  \bibinfo{author}{\bibfnamefont{J.}~\bibnamefont{{Ren}}},
  \bibinfo{author}{\bibfnamefont{J.-S.} \bibnamefont{{Wang}}},
  \bibnamefont{and} \bibinfo{author}{\bibfnamefont{B.}~\bibnamefont{{Li}}},
  \bibinfo{journal}{Phys. Rev. B} \textbf{\bibinfo{volume}{87}},
  \bibinfo{eid}{144101} (\bibinfo{year}{2013}).

\bibitem[{\citenamefont{{Lee} et~al.}(2015)\citenamefont{{Lee}, {Han}, and
  {Lee}}}]{Lee.Han.ea:PRB2015}
\bibinfo{author}{\bibfnamefont{H.}~\bibnamefont{{Lee}}},
  \bibinfo{author}{\bibfnamefont{J.~H.} \bibnamefont{{Han}}}, \bibnamefont{and}
  \bibinfo{author}{\bibfnamefont{P.~A.} \bibnamefont{{Lee}}},
  \bibinfo{journal}{Phys. Rev. B} \textbf{\bibinfo{volume}{91}},
  \bibinfo{eid}{125413} (\bibinfo{year}{2015}).

\bibitem[{\citenamefont{Hirschberger et~al.}(2015)\citenamefont{Hirschberger,
  Chisnell, Lee, and Ong}}]{Hirschberger.Chisnell.ea:PRL2015}
\bibinfo{author}{\bibfnamefont{M.}~\bibnamefont{Hirschberger}},
  \bibinfo{author}{\bibfnamefont{R.}~\bibnamefont{Chisnell}},
  \bibinfo{author}{\bibfnamefont{Y.~S.} \bibnamefont{Lee}}, \bibnamefont{and}
  \bibinfo{author}{\bibfnamefont{N.~P.} \bibnamefont{Ong}},
  \bibinfo{journal}{Phys. Rev. Lett.} \textbf{\bibinfo{volume}{115}},
  \bibinfo{pages}{106603} (\bibinfo{year}{2015}).

\bibitem[{\citenamefont{{Shindou}
  et~al.}(2013{\natexlab{a}})\citenamefont{{Shindou}, {Matsumoto}, {Murakami},
  and {Ohe}}}]{Shindou.Matsumoto.ea:PRB2013}
\bibinfo{author}{\bibfnamefont{R.}~\bibnamefont{{Shindou}}},
  \bibinfo{author}{\bibfnamefont{R.}~\bibnamefont{{Matsumoto}}},
  \bibinfo{author}{\bibfnamefont{S.}~\bibnamefont{{Murakami}}},
  \bibnamefont{and} \bibinfo{author}{\bibfnamefont{J.-i.} \bibnamefont{{Ohe}}},
  \bibinfo{journal}{Phys. Rev. B} \textbf{\bibinfo{volume}{87}},
  \bibinfo{eid}{174427} (\bibinfo{year}{2013}{\natexlab{a}}).

\bibitem[{\citenamefont{{Shindou}
  et~al.}(2013{\natexlab{b}})\citenamefont{{Shindou}, {Ohe}, {Matsumoto},
  {Murakami}, and {Saitoh}}}]{Shindou.Ohe.ea:PRB2013}
\bibinfo{author}{\bibfnamefont{R.}~\bibnamefont{{Shindou}}},
  \bibinfo{author}{\bibfnamefont{J.-i.} \bibnamefont{{Ohe}}},
  \bibinfo{author}{\bibfnamefont{R.}~\bibnamefont{{Matsumoto}}},
  \bibinfo{author}{\bibfnamefont{S.}~\bibnamefont{{Murakami}}},
  \bibnamefont{and} \bibinfo{author}{\bibfnamefont{E.}~\bibnamefont{{Saitoh}}},
  \bibinfo{journal}{Phys. Rev. B} \textbf{\bibinfo{volume}{87}},
  \bibinfo{eid}{174402} (\bibinfo{year}{2013}{\natexlab{b}}).

\bibitem[{\citenamefont{Mook et~al.}(2014)\citenamefont{Mook, Henk, and
  Mertig}}]{Mook.Henk.ea:PRB2014}
\bibinfo{author}{\bibfnamefont{A.}~\bibnamefont{Mook}},
  \bibinfo{author}{\bibfnamefont{J.}~\bibnamefont{Henk}}, \bibnamefont{and}
  \bibinfo{author}{\bibfnamefont{I.}~\bibnamefont{Mertig}},
  \bibinfo{journal}{Phys. Rev. B} \textbf{\bibinfo{volume}{90}},
  \bibinfo{pages}{024412} (\bibinfo{year}{2014}).

\bibitem[{\citenamefont{{Mook} et~al.}(2014)\citenamefont{{Mook}, {Henk}, and
  {Mertig}}}]{Mook.Henk.ea:PRB2014a}
\bibinfo{author}{\bibfnamefont{A.}~\bibnamefont{{Mook}}},
  \bibinfo{author}{\bibfnamefont{J.}~\bibnamefont{{Henk}}}, \bibnamefont{and}
  \bibinfo{author}{\bibfnamefont{I.}~\bibnamefont{{Mertig}}},
  \bibinfo{journal}{Phys. Rev. B} \textbf{\bibinfo{volume}{89}},
  \bibinfo{eid}{134409} (\bibinfo{year}{2014}).

\bibitem[{\citenamefont{Kovalev and Zyuzin}(2016)}]{Kovalev.Zyuzin:PRB2016}
\bibinfo{author}{\bibfnamefont{A.~A.} \bibnamefont{Kovalev}} \bibnamefont{and}
  \bibinfo{author}{\bibfnamefont{V.}~\bibnamefont{Zyuzin}},
  \bibinfo{journal}{Phys. Rev. B} \textbf{\bibinfo{volume}{93}},
  \bibinfo{pages}{161106} (\bibinfo{year}{2016}).

\bibitem[{\citenamefont{Fransson et~al.}(2016)\citenamefont{Fransson,
  Black-Schaffer, and Balatsky}}]{Fransson.ea:PRB16}
\bibinfo{author}{\bibfnamefont{J.}~\bibnamefont{Fransson}},
  \bibinfo{author}{\bibfnamefont{A.~M.} \bibnamefont{Black-Schaffer}},
  \bibnamefont{and} \bibinfo{author}{\bibfnamefont{A.~V.}
  \bibnamefont{Balatsky}}, \bibinfo{journal}{Phys. Rev. B}
  \textbf{\bibinfo{volume}{94}}, \bibinfo{pages}{075401}
  (\bibinfo{year}{2016}).

\bibitem[{\citenamefont{Owerre}(2016)}]{Owerre:apa2016}
\bibinfo{author}{\bibfnamefont{S.~A.} \bibnamefont{Owerre}},
  \bibinfo{journal}{Journal of Applied Physics} \textbf{\bibinfo{volume}{120}},
  \bibinfo{eid}{043903} (\bibinfo{year}{2016}).

\bibitem[{\citenamefont{Kim et~al.}(2016)\citenamefont{Kim, Ochoa, Zarzuela,
  and Tserkovnyak}}]{Kim.Ochoa.ea:apa2016}
\bibinfo{author}{\bibfnamefont{S.~K.} \bibnamefont{Kim}},
  \bibinfo{author}{\bibfnamefont{H.}~\bibnamefont{Ochoa}},
  \bibinfo{author}{\bibfnamefont{R.}~\bibnamefont{Zarzuela}}, \bibnamefont{and}
  \bibinfo{author}{\bibfnamefont{Y.}~\bibnamefont{Tserkovnyak}},
  \bibinfo{journal}{arXiv preprint arXiv:1603.04827}  (\bibinfo{year}{2016}).

\bibitem[{\citenamefont{Jungwirth et~al.}(2016)\citenamefont{Jungwirth, Marti,
  Wadley, and Wunderlich}}]{Jungwirth.Marti.ea:NN2016}
\bibinfo{author}{\bibfnamefont{T.}~\bibnamefont{Jungwirth}},
  \bibinfo{author}{\bibfnamefont{X.}~\bibnamefont{Marti}},
  \bibinfo{author}{\bibfnamefont{P.}~\bibnamefont{Wadley}}, \bibnamefont{and}
  \bibinfo{author}{\bibfnamefont{J.}~\bibnamefont{Wunderlich}},
  \bibinfo{journal}{Nat. Nanotechnol.} \textbf{\bibinfo{volume}{11}},
  \bibinfo{pages}{231} (\bibinfo{year}{2016}).

\bibitem[{\citenamefont{{Ohnuma} et~al.}(2013)\citenamefont{{Ohnuma}, {Adachi},
  {Saitoh}, and {Maekawa}}}]{Ohnuma.Adachi.ea:PRB2013}
\bibinfo{author}{\bibfnamefont{Y.}~\bibnamefont{{Ohnuma}}},
  \bibinfo{author}{\bibfnamefont{H.}~\bibnamefont{{Adachi}}},
  \bibinfo{author}{\bibfnamefont{E.}~\bibnamefont{{Saitoh}}}, \bibnamefont{and}
  \bibinfo{author}{\bibfnamefont{S.}~\bibnamefont{{Maekawa}}},
  \bibinfo{journal}{Phys. Rev. B} \textbf{\bibinfo{volume}{87}},
  \bibinfo{eid}{014423} (\bibinfo{year}{2013}).

\bibitem[{\citenamefont{{Rezende} et~al.}(2016)\citenamefont{{Rezende},
  {Rodr{\'{\i}}guez-Su{\'a}rez}, and
  {Azevedo}}}]{Rezende.Rodriguez-Suarez.ea:PRB2016}
\bibinfo{author}{\bibfnamefont{S.~M.} \bibnamefont{{Rezende}}},
  \bibinfo{author}{\bibfnamefont{R.~L.}
  \bibnamefont{{Rodr{\'{\i}}guez-Su{\'a}rez}}}, \bibnamefont{and}
  \bibinfo{author}{\bibfnamefont{A.}~\bibnamefont{{Azevedo}}},
  \bibinfo{journal}{Phys. Rev. B} \textbf{\bibinfo{volume}{93}},
  \bibinfo{eid}{014425} (\bibinfo{year}{2016}).

\bibitem[{\citenamefont{{Matsumoto} et~al.}(2014)\citenamefont{{Matsumoto},
  {Shindou}, and {Murakami}}}]{Matsumoto.Shindou.ea:PRB2014}
\bibinfo{author}{\bibfnamefont{R.}~\bibnamefont{{Matsumoto}}},
  \bibinfo{author}{\bibfnamefont{R.}~\bibnamefont{{Shindou}}},
  \bibnamefont{and}
  \bibinfo{author}{\bibfnamefont{S.}~\bibnamefont{{Murakami}}},
  \bibinfo{journal}{Phys. Rev. B} \textbf{\bibinfo{volume}{89}},
  \bibinfo{eid}{054420} (\bibinfo{year}{2014}).

\bibitem[{\citenamefont{{Luttinger}}(1964)}]{Luttinger:PR1964}
\bibinfo{author}{\bibfnamefont{J.~M.} \bibnamefont{{Luttinger}}},
  \bibinfo{journal}{Phys. Rev.} \textbf{\bibinfo{volume}{135}},
  \bibinfo{pages}{1505} (\bibinfo{year}{1964}).

\bibitem[{\citenamefont{{Tatara}}(2015)}]{Tatara:PRB2015}
\bibinfo{author}{\bibfnamefont{G.}~\bibnamefont{{Tatara}}},
  \bibinfo{journal}{Phys. Rev. B} \textbf{\bibinfo{volume}{92}},
  \bibinfo{eid}{064405} (\bibinfo{year}{2015}).

\bibitem[{\citenamefont{{Tsirlin} et~al.}(2010)\citenamefont{{Tsirlin},
  {Janson}, and {Rosner}}}]{Tsirlin.Janson.ea:PRB2010}
\bibinfo{author}{\bibfnamefont{A.~A.} \bibnamefont{{Tsirlin}}},
  \bibinfo{author}{\bibfnamefont{O.}~\bibnamefont{{Janson}}}, \bibnamefont{and}
  \bibinfo{author}{\bibfnamefont{H.}~\bibnamefont{{Rosner}}},
  \bibinfo{journal}{Phys. Rev. B} \textbf{\bibinfo{volume}{82}},
  \bibinfo{eid}{144416} (\bibinfo{year}{2010}).

\bibitem[{\citenamefont{{Liu} et~al.}(2011)\citenamefont{{Liu}, {Berlijn},
  {Yin}, {Ku}, {Tsvelik}, {Kim}, {Gretarsson}, {Singh}, {Gegenwart}, and
  {Hill}}}]{Liu.Berlijn.ea:PRB2011}
\bibinfo{author}{\bibfnamefont{X.}~\bibnamefont{{Liu}}},
  \bibinfo{author}{\bibfnamefont{T.}~\bibnamefont{{Berlijn}}},
  \bibinfo{author}{\bibfnamefont{W.-G.} \bibnamefont{{Yin}}},
  \bibinfo{author}{\bibfnamefont{W.}~\bibnamefont{{Ku}}},
  \bibinfo{author}{\bibfnamefont{A.}~\bibnamefont{{Tsvelik}}},
  \bibinfo{author}{\bibfnamefont{Y.-J.} \bibnamefont{{Kim}}},
  \bibinfo{author}{\bibfnamefont{H.}~\bibnamefont{{Gretarsson}}},
  \bibinfo{author}{\bibfnamefont{Y.}~\bibnamefont{{Singh}}},
  \bibinfo{author}{\bibfnamefont{P.}~\bibnamefont{{Gegenwart}}},
  \bibnamefont{and} \bibinfo{author}{\bibfnamefont{J.~P.}
  \bibnamefont{{Hill}}}, \bibinfo{journal}{Phys. Rev. B}
  \textbf{\bibinfo{volume}{83}}, \bibinfo{eid}{220403} (\bibinfo{year}{2011}).

\bibitem[{\citenamefont{{Lee} et~al.}(2012)\citenamefont{{Lee}, {Choi}, {Kim},
  {Sim}, {Won}, {Lee}, {Kim}, {Hur}, and {Park}}}]{Lee.Choi.ea:JoPCM2012}
\bibinfo{author}{\bibfnamefont{S.}~\bibnamefont{{Lee}}},
  \bibinfo{author}{\bibfnamefont{S.}~\bibnamefont{{Choi}}},
  \bibinfo{author}{\bibfnamefont{J.}~\bibnamefont{{Kim}}},
  \bibinfo{author}{\bibfnamefont{H.}~\bibnamefont{{Sim}}},
  \bibinfo{author}{\bibfnamefont{C.}~\bibnamefont{{Won}}},
  \bibinfo{author}{\bibfnamefont{S.}~\bibnamefont{{Lee}}},
  \bibinfo{author}{\bibfnamefont{S.~A.} \bibnamefont{{Kim}}},
  \bibinfo{author}{\bibfnamefont{N.}~\bibnamefont{{Hur}}}, \bibnamefont{and}
  \bibinfo{author}{\bibfnamefont{J.-G.} \bibnamefont{{Park}}},
  \bibinfo{journal}{J. Phys.: Condens. Matter.} \textbf{\bibinfo{volume}{24}},
  \bibinfo{eid}{456004} (\bibinfo{year}{2012}).

\bibitem[{\citenamefont{{Singh} et~al.}(2012)\citenamefont{{Singh}, {Manni},
  {Reuther}, {Berlijn}, {Thomale}, {Ku}, {Trebst}, and
  {Gegenwart}}}]{Singh.Manni.ea:PRL2012}
\bibinfo{author}{\bibfnamefont{Y.}~\bibnamefont{{Singh}}},
  \bibinfo{author}{\bibfnamefont{S.}~\bibnamefont{{Manni}}},
  \bibinfo{author}{\bibfnamefont{J.}~\bibnamefont{{Reuther}}},
  \bibinfo{author}{\bibfnamefont{T.}~\bibnamefont{{Berlijn}}},
  \bibinfo{author}{\bibfnamefont{R.}~\bibnamefont{{Thomale}}},
  \bibinfo{author}{\bibfnamefont{W.}~\bibnamefont{{Ku}}},
  \bibinfo{author}{\bibfnamefont{S.}~\bibnamefont{{Trebst}}}, \bibnamefont{and}
  \bibinfo{author}{\bibfnamefont{P.}~\bibnamefont{{Gegenwart}}},
  \bibinfo{journal}{Phys. Rev. Lett.} \textbf{\bibinfo{volume}{108}},
  \bibinfo{eid}{127203} (\bibinfo{year}{2012}).

\bibitem[{\citenamefont{{Choi} et~al.}(2012)\citenamefont{{Choi}, {Coldea},
  {Kolmogorov}, {Lancaster}, {Mazin}, {Blundell}, {Radaelli}, {Singh},
  {Gegenwart}, {Choi} et~al.}}]{Choi.Coldea.ea:PRL2012}
\bibinfo{author}{\bibfnamefont{S.~K.} \bibnamefont{{Choi}}},
  \bibinfo{author}{\bibfnamefont{R.}~\bibnamefont{{Coldea}}},
  \bibinfo{author}{\bibfnamefont{A.~N.} \bibnamefont{{Kolmogorov}}},
  \bibinfo{author}{\bibfnamefont{T.}~\bibnamefont{{Lancaster}}},
  \bibinfo{author}{\bibfnamefont{I.~I.} \bibnamefont{{Mazin}}},
  \bibinfo{author}{\bibfnamefont{S.~J.} \bibnamefont{{Blundell}}},
  \bibinfo{author}{\bibfnamefont{P.~G.} \bibnamefont{{Radaelli}}},
  \bibinfo{author}{\bibfnamefont{Y.}~\bibnamefont{{Singh}}},
  \bibinfo{author}{\bibfnamefont{P.}~\bibnamefont{{Gegenwart}}},
  \bibinfo{author}{\bibfnamefont{K.~R.} \bibnamefont{{Choi}}},
  \bibnamefont{et~al.}, \bibinfo{journal}{Phys. Rev. Lett.}
  \textbf{\bibinfo{volume}{108}}, \bibinfo{eid}{127204} (\bibinfo{year}{2012}).

\bibitem[{\citenamefont{Auerbach}(1994)}]{Auerbach:GTiCP1994}
\bibinfo{author}{\bibfnamefont{A.}~\bibnamefont{Auerbach}},
  \emph{\bibinfo{title}{Interacting Electrons and Quantum Magnetism}}
  (\bibinfo{publisher}{Springer New York}, \bibinfo{year}{1994}).

\bibitem[{\citenamefont{{Smrcka} and {Streda}}(1977)}]{Smrcka.Streda:JPC1977}
\bibinfo{author}{\bibfnamefont{L.}~\bibnamefont{{Smrcka}}} \bibnamefont{and}
  \bibinfo{author}{\bibfnamefont{P.}~\bibnamefont{{Streda}}},
  \bibinfo{journal}{J. Phys. C} \textbf{\bibinfo{volume}{10}},
  \bibinfo{pages}{2153} (\bibinfo{year}{1977}).

\bibitem[{\citenamefont{Zyuzin and Kovalev}(2016)}]{Zyuzin.Kovalev.MM2016}
\bibinfo{author}{\bibfnamefont{V.~A.} \bibnamefont{Zyuzin}} \bibnamefont{and}
  \bibinfo{author}{\bibfnamefont{A.~A.} \bibnamefont{Kovalev}},
  \bibinfo{journal}{Bulletin of the American Physical Society, 2016 March
  Meeting} \textbf{\bibinfo{volume}{61}}, \bibinfo{pages}{B6.00005}
  (\bibinfo{year}{2016}).

\bibitem[{\citenamefont{Cheng et~al.}(2016)\citenamefont{Cheng, Okamoto, and
  Xiao}}]{Cheng.Okamoto.Xiao:ArXiv}
\bibinfo{author}{\bibfnamefont{R.}~\bibnamefont{Cheng}},
  \bibinfo{author}{\bibfnamefont{S.}~\bibnamefont{Okamoto}}, \bibnamefont{and}
  \bibinfo{author}{\bibfnamefont{D.}~\bibnamefont{Xiao}},
  \bibinfo{journal}{arXiv:1606.01952}  (\bibinfo{year}{2016}).

\end{thebibliography}

\newpage
\begin{widetext}
\appendix
\begin{center}
{{\bf Supplemental material}}
\end{center}

In this supplementary a letter $\beta$ will note two different quantities, namely direction of the temperature gradient, and inverse temperature $\beta= 1/T$ when talking about the Bose-Einstein distribution function. Letter $\mathrm{T}$ will stand for temperature, transpose symbol, and a paraunitary matrix $T_{\bf k}$.

\section{A model of antiferromagnet}
We adopt the Luttinger formalism Ref.\onlinecite{Luttinger:PR1964} to study the response of the system to the temperature gradient. 
Define a Hamiltonian corresponding to a boson system with anomalous terms
\begin{align}
H = \frac{1}{2} \int d{\bf r}{\tilde \Psi}^{\dag}({\bf r}) {\hat H}({\bf r}) {\tilde \Psi}({\bf r}), 
\end{align}
where ${\tilde \Psi}({\bf r}) = \left( 1+\frac{{\bf r}{\bm \nabla}\chi}{2} \right)\Psi({\bf r}) \equiv \xi({\bf r})\Psi({\bf r}) $ with ${{\bm \nabla}\chi}$ being the temperature gradient, the boson operators are $\Psi^{\dag}({\bf r}) = [\nu^{\dag}_{1}({\bf r}),...,\nu^{\dag}_{N}({\bf r}), \nu_{1}({\bf r}),...,\nu_{N}({\bf r}) ]$, with commutation relations $[\nu_{i}({\bf r}),\nu^{\dag}_{j}({\bf r}^{\prime}) ] = \delta_{i,j}\delta_{{\bf r},{\bf r}^{\prime}}$. The commutation relations are
\begin{align}
&
[\Psi_{i}({\bf r}),\Psi^{\dag}_{j}({\bf r}^{\prime})] = \left( \sigma_{3} \right)_{ij}\delta_{{\bf r},{\bf r}^\prime}
\\
&
[\Psi^{\dag}_{i}({\bf r}),\Psi^{\dag}_{j}({\bf r}^{\prime})] =-i \left( \sigma_{2} \right)_{ij}\delta_{{\bf r},{\bf r}^\prime}
\\
&
[\Psi_{i}({\bf r}),\Psi_{j}({\bf r}^{\prime})] = i \left( \sigma_{2} \right)_{ij}\delta_{{\bf r},{\bf r}^\prime},
\end{align}
where 
\begin{align}
\sigma_{1} = \left[ \begin{array}{cc}
0 & 1_{N \times N}  \\
1_{N \times N} & 0   \end{array}  \right], 
~~
\sigma_{2} = \left[ \begin{array}{cc}
0 & -i1_{N \times N}  \\
i1_{N \times N} & 0   \end{array}  \right],
~~
\sigma_{3} = \left[ \begin{array}{cc}
1_{N \times N} & 0  \\
0 & -1_{N \times N}    \end{array}  \right],
\end{align}
are Pauli matrices acting on the extended space.

\subsection{Diagonal basis}
The Hamiltonian is diagonalized with a help of a matrix $T_{{\bf k}}$, such that 
\begin{align}
 \varepsilon_{{\bf k}} = T_{{\bf k}}^{\dag} {\hat H} T_{{\bf k}} = 
\left[ \begin{array}{cc} E_{{\bf k}} & 0 \\ 0 & E_{-{\bf k}}  \end{array} \right],
\end{align}
where $T_{\bf k}^{\dag}$ is a paraunitary matrix, obeying
\begin{align}
T_{\bf k}^{\dag} \sigma_{3}T_{\bf k} = \sigma_{3}.
\end{align}

It is convenient to present a boson operator in terms of the modes which correspond to the diagonalized form of the Hamiltonian. In this way, in normal modes
\begin{align}
\Psi_{{\bf k}}^{\dag} = \Gamma_{{\bf k}}^{\dag} T_{{\bf k}}^{\dag},
\end{align}
where
\begin{align}
\Gamma_{\bf k} = \left[ \begin{array}{c}
{\hat \gamma}_{{\bf k}} \\
 {\hat \gamma}^{\dag}_{-{\bf k}}
\end{array} \right].
\end{align} 
In normal modes the Hamiltonian becomes
\begin{align}
\Psi^{\dag}_{\bf k}{\hat H}_{\bf k}\Psi_{\bf k} = \Gamma_{\bf k}^{\dag}\varepsilon_{\bf k}\Gamma_{\bf k}.
\end{align}

It is important to derive an identity between $T_{\bf k}$ and $T_{-{\bf k}}$ matrices.
By applying a particle-hole symmetry transformation, namely
\begin{align}
{\hat H}_{\bf k} = \sigma_{1}\left( {\hat H}^{\mathrm{T}}_{-{\bf k}}\right)\sigma_{1},
\end{align}
to the eigenvalue problem for $T_{\bf k}$ 
\begin{align}
& 
{\hat H}_{\bf k}T_{\bf k} =\sigma_{3}T_{\bf k}\sigma_{3}\varepsilon_{\bf k},
\\
& 
 T_{\bf k}^{\dag}{\hat H}_{\bf k} = \varepsilon_{\bf k}\sigma_{3}T_{\bf k}^{\dag}\sigma_{3},
\end{align}
written in two equivalent ways, we obtain
\begin{align}
\left( \sigma_{1}T_{-{\bf k}}^{\mathrm{T}}\sigma_{1} \right) {\hat H}_{\bf k} = \varepsilon_{\bf k}\sigma_{3}\left( \sigma_{1}T_{-{\bf k}}^{\mathrm{T}}\sigma_{1} \right)\sigma_{3}.
\end{align}
From where we can deduce
\begin{align}
T_{\bf k}^{\dag} = P_{\bf k}^{\dag} \left( \sigma_{1}T_{-{\bf k}}^{\mathrm{T}}\sigma_{1} \right) ,
\end{align}
where $P_{\bf k}$ is a matrix obeying  a paraunitarity condition 
\begin{align}
P^{\dag}_{\bf k}\sigma_{3}P_{\bf k} = \sigma_{3}.
\end{align}
Another condition that can be deduced is
\begin{align}
P_{\bf k}^{\dag}\sigma_{3}\varepsilon_{\bf k} = \varepsilon_{\bf k}\sigma_{3}P_{\bf k}^{\dag}.
\end{align}
Since $\varepsilon_{\bf k}$ is diagonal, one can show $P_{\bf k}$ is a diagonal matrix with phase factors elements, i.e. $\left( P_{\bf k}\right)_{nn} = e^{i\theta_{{\bf k}n}}$. 
From it we can conclude another identity, namely
\begin{align}
P_{\bf k}^{\dag}P_{\bf k} = 1.
\end{align}

\subsection{Holstein-Primakoff bosons}
Here we review a transformation from spins to bosons, called Holstein-Primakoff transformation (for a review Ref. \onlinecite{Auerbach:GTiCP1994} ).  
For a given spin ${\bf S}$ described by a classical direction ${\bm \Omega}$, one introduces the basis vectors $({\bf e}^{(1)},{\bf e}^{(2)},{\bm \Omega})$ such as
\begin{align}
{\bf e}^{(1)}\times {\bf e}^{(2)} = {\bm \Omega}.
\end{align}
Lowering and raising operators in the reference frame of the spin is then
\begin{align}
S^{\pm} = {\bf S}{\bf e}^{(1)} \pm i {\bf S}{\bf e}^{(2)}.
\end{align}
We then  introduce Holstein-Primakoff bosons $c$ and $c^{\dag}$ as
\begin{align}
S^{+} = \left(\sqrt{2S-c^{\dag}c} \right) c , ~~ S^{-} =  c^{\dag} \left( \sqrt{2S-c^{\dag}c} \right), ~~ {\bf S}{\bm \Omega} = S - c^{\dag}c.
\end{align}
Let us take two spins ${\bf S}_{\mathrm{A}}$ and ${\bf S}_{\mathrm{B}}$, and assume that there is an arbitrary angle between them.  
Introduce rotation matrices
\begin{align}
R_{\theta} = 
\left[
\begin{array}{ccc}
\cos(\theta) & 0 & -\sin(\theta) \\
0 & 1 & 0 \\
\sin(\theta) & 0 & \cos(\theta)
\end{array}
\right],
~~ 
R_{\phi} = 
\left[
\begin{array}{ccc}
\cos(\phi) & \sin(\phi) & 0\\
-\sin(\phi) & \cos(\phi) & 0 \\
0 & 0 & 1
\end{array}
\right],
\end{align}
such that ${\bf S}_{\mathrm{B}} = R_{\phi}R_{\theta} {\bf S}^{[z]}_{\mathrm{B}}$, where ${\bf S}^{[z]}_{\mathrm{B}}$ points in $z-$direction. 
Holstein-Primakoff transformation
\begin{align}
S_{\mathrm{A}} = \left[
\begin{array}{c}
\frac{1}{2}(S_{\mathrm{A}}^{+}+S_{\mathrm{A}}^{-}) \\
\frac{1}{2i}(S_{\mathrm{A}}^{+}-S_{\mathrm{A}}^{-}) \\
S_{\mathrm{A}} -  a^{\dag}a
\end{array}
\right],
~~
S_{\mathrm{B}} = R_{\phi}R_{\theta}\left[
\begin{array}{c}
\frac{1}{2}(S_{\mathrm{B}}^{+}+S_{\mathrm{B}}^{-}) \\
\frac{1}{2i}(S_{\mathrm{B}}^{+}-S_{\mathrm{B}}^{-}) \\
S_{\mathrm{B}} -  b^{\dag}b
\end{array}
\right].
\end{align}

As an example, consider a Neel order with two sublatttices A and B such that ${\bm \Omega}_{\mathrm{A}/\mathrm{B}} = \pm {\bf e}^{z}$.  
We can write for the A sublattice   
\begin{align}
S^{\pm} = S^{x} \pm iS^{y}, ~~ S^{z} = S-a^{\dag}a ,
\end{align}
and for B sublattice
\begin{align}
S^{\pm} = -S^{x} \pm iS^{y}, ~~ S^{z} = -S + b^{\dag}b.
\end{align}
Then the exchange Hamiltonian becomes
\begin{align}
{\bf S}_{\mathrm{A}}{\bf S}_{\mathrm{B}} = 
-\frac{2}{2}\left( S_{\mathrm{A}}^{+}S_{\mathrm{B}}^{+} + S_{\mathrm{A}}^{-}S_{\mathrm{B}}^{-}  \right) +  S_{\mathrm{A}}^{z}S_{\mathrm{B}}^{z}
=-S(ab+  a^{\dag}b^{\dag}) + S(a^{\dag}a + b^{\dag}b)
\end{align}
Dzyaloshinskii-Moriya interaction 
\begin{align}
\left[ {\bf S}_{\mathrm{A}}\times {\bf S}_{\mathrm{B}} \right]_{z} = \frac{1}{2i}\left( S_{\mathrm{A}}^{+}S_{\mathrm{B}}^{+} - S_{\mathrm{A}}^{-}S_{\mathrm{B}}^{-} \right)
= -iS(ab - a^{\dag}b^{\dag})
\end{align}
\begin{align}
\left[ {\bf S}_{\mathrm{A}1}\times {\bf S}_{\mathrm{A}2} \right]_{z} = \frac{1}{2i}\left(  S_{\mathrm{A}1}^{-}S_{\mathrm{A}2}^{+} - S_{\mathrm{A}1}^{+}S_{\mathrm{A}2}^{-} \right) = -iS(a_{1}^{\dag}a_{2} - a_{2}^{\dag}a_{1})
\end{align}
\begin{align}
\left[ {\bf S}_{\mathrm{B}1}\times {\bf S}_{\mathrm{B}2} \right]_{z} = \frac{1}{2i}\left(  S_{\mathrm{B}1}^{-}S_{\mathrm{B}2}^{+} - S_{\mathrm{B}1}^{+}S_{\mathrm{B}2}^{-} \right) = iS(b_{1}^{\dag}b_{2} - b_{2}^{\dag}b_{1})
\end{align}

\subsection{Current}
In deriving the following identities, we assume a system to be on a lattice, such that the Hamiltonian density acts on the operators as ${\hat H}\Psi({\bf r})=\sum_{{\bm \delta}} H_{{\bm \delta}}\Psi({\bf r}+{\bm \delta})$, where ${\bm \delta}$ is a distance between sites on the lattice.

Let us define an arbitrary operator ${\hat O}$, and demand from this ${\hat O}$ to not have translation operators in its definition, i.e. it should commute with the position operator.  
The density of such an operator is 
\begin{align}
{\cal O}({\bf r}) = \frac{1}{2}\Psi^{\dag}({\bf r}){\hat O}\Psi({\bf r}) ,
\end{align}
which can, for example, correspond to density of magnons or spin density.  
Let us then calculate the time evolution of this operator
\begin{align}
\frac{\partial {\cal O}({\bf r})}{\partial t} = i[ H,{\cal O}({\bf r})].
\end{align}
The commutator written in the band index 
\begin{align}
[H,{\cal O}({\bf r})] 
&
= -\frac{1}{4}\sum_{\bm \delta}\int d{\bf r}^{\prime}\left[  \Psi_{n}^{\dag}({\bf r}){\hat O}_{nn^\prime}\Psi_{n^\prime}({\bf r}) {\tilde \Psi}_{m}^{\dag}({\bf r}^\prime) 
(H_{{\bm \delta}})_{mk}{\tilde \Psi}_{k}({\bf r}^\prime + {\bm \delta})
-
 {\tilde \Psi}_{m}^{\dag}({\bf r}^\prime) 
(H_{{\bm \delta}})_{mk}{\tilde \Psi}_{k}({\bf r}^\prime + {\bm \delta})\Psi_{m}^{\dag}({\bf r}){\hat O}_{nn^\prime}\Psi_{n^\prime}({\bf r})  \right]
\nonumber
\\
&
= - \frac{1}{2}\sum_{\bm \delta} \left\{ \Psi_{n}^{\dag}({\bf r}) {\hat O}_{nn^\prime}\left( \sigma_{3} \right)_{n^\prime m}
[ \xi({\bf r})  H_{{\bm \delta}}  \xi({\bf r} + {\bm \delta}) ]_{mk}
\Psi_{k}({\bf r} + {\bm \delta})
-
 \Psi_{m}^{\dag}({\bf r} - {\bm \delta}) [\xi({\bf r}- {\bm \delta}) 
H_{{\bm \delta}}\xi({\bf r})]_{mk} (\sigma_{3})_{kn} {\hat O}_{nn^\prime}\Psi_{n^\prime}({\bf r})
\right\}
\nonumber
\\
&
=- \frac{1}{2} \sum_{\bm \delta}\left[ {\tilde \Psi}^{\dag}({\bf r}) {\hat O}\sigma_{3}H_{{\bm \delta}} {\tilde \Psi}({\bf r} + {\bm \delta})
- {\tilde \Psi}^{\dag}({\bf r} - {\bm \delta}) H_{{\bm \delta}}\sigma_{3}{\hat O} {\tilde \Psi}({\bf r}) \right]
\nonumber
\\
&
= i \frac{1}{2}{\bm \nabla} {\tilde \Psi}^{\dag}({\bf r})\left({\hat {\bf v}}\sigma_{3}{\hat O} + {\hat O}\sigma_{3}{\hat {\bf v}} \right){\tilde \Psi}({\bf r})
- \frac{1}{2}{\tilde \Psi}^{\dag}({\bf r})\left( {\hat O}\sigma_{3}{\hat H} - {\hat H}\sigma_{3}{\hat O}  \right){\tilde \Psi}({\bf r}) ,
\end{align}
we observe that for the current of the operator ${\hat O}$ to be well defined, condition ${\hat O}\sigma_{3}{\hat H} -{\hat H}\sigma_{3}{\hat O}  = 0$ must be satisfied by the operator.
We defined velocity as ${\bf v} = i \sum_{{\bm \delta}}{\bm \delta} H_{\bm \delta} e^{i{\hat {\bf p}}{\bm \delta}} = i\left[ H,{\bf r}\right]$.
 We can use the mentioned above commutation to show that ${\hat {\bf v}}\sigma_{3}{\hat O} = {\hat O}\sigma_{3}{\hat {\bf v}}$.
 The current then will be a well defined quantity, equal to 
\begin{align}
{\bf j}_{O} =  {\tilde \Psi}^{\dag}({\bf r})  {\hat O}\sigma_{3}{\hat {\bf v}}  {\tilde \Psi}({\bf r}).
\end{align} 
We will show that for Neel order operator ${\hat O}$ might correspond to spin density.

\section{Response to temperature gradient}
Let us assume we defined such an operator ${\hat O}$ that satisfies the condition. We split the current in to two parts
\begin{align}
{\bf j}_{\mathrm{O}}^{[0]} =   \Psi^{\dag}({\bf r})  {\hat O}\sigma_{3}{\hat {\bf v}}   \Psi({\bf r}),
\end{align}
and
\begin{align}
{\bf j}_{\mathrm{O}}^{[1]} = \frac{1}{2} \Psi^{\dag}({\bf r}){\hat O}\sigma_{3}
\left( r_{\beta} {\hat {\bf v}} + {\hat {\bf v}} r_{\beta} \right)\Psi({\bf r}) \nabla_{\beta} \chi .
\end{align}
In the following we will omit the $\mathrm{O}$ index from the current for the sake of simplicity. 
We will be working with the macroscopic quantities, such as ${\bf J}^{[0]} = \frac{1}{V}\int d{\bf r} {\bf j}^{[0]}({\bf r})$ and ${\bf J}^{[1]} = \frac{1}{V}\int d{\bf r} {\bf j}^{[1]}({\bf r})$, where $V$ is the volume of the system.
When calculating the current we need to consider 
\begin{align}
J_{\alpha} = \left< J_{\alpha}^{[0]}\right>_{\mathrm{ne}} +  \left< J_{\alpha}^{[1]}\right>_{\mathrm{eq}},
\end{align}
currents. The first current, $\left< J_{\alpha}^{[0]}\right>_{\mathrm{ne}}$, is estimated over the non-equilibrium states, whose evolution operator is defined by the perturbing Hamiltonian. It's expression will be derived via Kubo formula. 
The second current, $\left< J_{\alpha}^{[1]}\right>_{\mathrm{eq}}$, is estimated over the equilibrium states. We will refer to this current as magnetization driven.  

Time ordered averages over the equilibrium state of the system are performed via the following rules for the diagonal basis boson operators: 
\begin{align}
\left<T_{\tau} \gamma^{\dag}_{{\bf k} n}(\tau^\prime + \tau)\gamma_{{\bf k}^\prime m}(\tau^\prime)  \right> = \delta_{n,m}\delta_{{\bf k},{\bf k}^\prime}g[(\varepsilon_{\bf k})_{nn}]e^{\tau (\varepsilon_{\bf k})_{nn} },
\end{align}
\begin{align}
\left< T_{\tau} \gamma_{{\bf k}n}(\tau^\prime + \tau) \gamma^{\dag}_{{\bf k}^\prime m}(\tau^\prime)  \right> = - \delta_{n,m}\delta_{{\bf k},{\bf k}^\prime}g[-(\varepsilon_{\bf k})_{nn}]e^{-\tau (\varepsilon_{\bf k})_{nn}},
\end{align}
where $g(\epsilon) = (e^{\beta\epsilon} - 1)^{-1}$ is the Bose-Einstein distribution function with $\beta = 1/T$, and where for $\tau > 0$ the time ordering is already satisfied.
An identity 
\begin{align}
g(\epsilon)+1 = \frac{1}{e^{\beta \epsilon} - 1} + 1 = - \frac{1}{e^{-\beta \epsilon} - 1} = - g(-\epsilon)
\end{align}
was used in deriving the averages.

\subsection{Kubo formula}
An average of ${\bf J}^{[0]}$ over the non-equilibrium states can be conveniently captured with a help of Kubo formula. 
A goal of this section is to derive an expression for $S_{\alpha\beta}$, which enters the Kubo formula as follows.
We write for the currents $\alpha$ component
\begin{align}
\left<  J_{\alpha}^{[0]}  \right>_{\mathrm{ne}}
=  -\lim_{\omega \to 0} \frac{\partial}{\partial \omega} \int_{0}^{1/T} d\tau e^{i\omega \tau} \left<T_{\tau}J_{\alpha}^{[0]}(\tau)J_{\beta}^{\mathrm{Q}}(0)  \right>
 \equiv \frac{1}{V} S_{\alpha\beta} \nabla_{\beta}\chi ,
\end{align}
where $\omega = 2\pi n/T$ is boson Matsubara frequency, $V$ is the volume of the system, and where $J_{\beta}^{\mathrm{Q}}$ is a current defined as follows. We define Hamiltonian densities $h_{0} = \frac{1}{2}\Psi^{\dag}({\bf r}){\hat H}\Psi({\bf r})$ and   $h^{\prime}({\bf r}) = \frac{1}{2}\Psi^{\dag}({\bf r}) \left(  r_{\beta}{\hat H} + {\hat H}r_{\beta}   \right)\Psi({\bf r})\nabla_{\beta}\chi$, corresponding to unperturbed and perturbing Hamiltonians. 
The current which enters the Kubo formula is defined through a commutator
\begin{align}
\frac{\partial h^{\prime}({\bf r})}{\partial t} &= \frac{i}{\hbar}\int d{\bf r}^\prime \left[ h_{0}({\bf r}^\prime),h^\prime({\bf r})\right]  
\nonumber
\\
&
=  {\bm \nabla} \frac{1}{2}\left[\Psi^{\dag}({\bf r})\left( {\hat H}\sigma_{3}r_{\beta}{\hat {\bf v}}  + {\hat {\bf v}}r_{\beta}\sigma_{3}{\hat H}  \right)\Psi({\bf r}) \right] \nabla_{\beta}\chi
- \frac{1}{4}\Psi^{\dag}({\bf r})\left( {\hat H}\sigma_{3}{\hat v}_{\beta} + {\hat v}_{\beta}\sigma_{3}{\hat H} \right)\Psi({\bf r})\nabla_{\beta}\chi,
\end{align}
which we integrate over the space, and obtain 
\begin{align}
\int d{\bf r} \frac{\partial h^{\prime}({\bf r})}{\partial t} =  -
\frac{1}{4} \int d{\bf r} \Psi^{\dag}({\bf r})\left( {\hat H}\sigma_{3}{\hat v}_{\beta} + {\hat v}_{\beta}\sigma_{3} {\hat H} \right)\Psi({\bf r})\nabla_{\beta}\chi
 \equiv - \int d{\bf r} j_{\beta}^{\mathrm{Q}}\nabla_{\beta}\chi 
\equiv - J_{\beta}^{\mathrm{Q}}\nabla_{\beta}\chi.
\end{align}
We define $S_{\alpha\beta}$ as follows
\begin{align}
&
S_{\alpha\beta} = - \frac{1}{4} \lim_{\omega \to 0}\frac{\partial}{\partial \omega} \int_{0}^{1/T}d\tau e^{i\omega\tau}\sum_{{\bf k}{\bf k}^\prime} \left< \Psi^{\dag}_{{\bf k}}(\tau){\hat O}\sigma_{3} v_{\alpha {\bf k}} \Psi_{{\bf k}}(\tau) \Psi^{\dag}_{{\bf k}^\prime}(0)
\left[ H_{{\bf k}^\prime}\sigma_{3}v_{\beta {\bf k}^\prime} + v_{\beta {\bf k}^\prime}\sigma_{3}H_{{\bf k}^\prime} \right]\Psi_{{\bf k}^\prime}(0) \right>
\nonumber
\\
&= - \frac{1}{4} \lim_{\omega \to 0} \frac{\partial}{\partial \omega} \int_{0}^{1/T}d\tau e^{i\omega\tau} \left< \Gamma_{{\bf k}}^{\dag}(\tau) {\tilde V}_{\alpha{\bf k}}\Gamma_{{\bf k}}(\tau) \Gamma^{\dag}_{{\bf k}^\prime}(0)
\left( \varepsilon_{{\bf k}^\prime}\sigma_{3}{\tilde v}_{\beta {\bf k}^\prime} + {\tilde v}_{\beta {\bf k}^\prime}\sigma_{3}\varepsilon_{{\bf k}^\prime}  \right)\Gamma_{{\bf k}^\prime}(0) \right>,
\end{align}
where ${\tilde V}_{\alpha \bf k} = T_{\bf k}^{\dag}{\hat O}\sigma_{3}v_{\alpha \bf k} T_{\bf k}$.
The average of the boson operators over the equilibrium state of the system is
\begin{align}
&
\left< \Gamma^{\dag}_{{\bf k},n}(\tau)\Gamma_{{\bf k},m}(\tau)\Gamma^{\dag}_{{\bf k}^\prime ,t}(0)\Gamma_{{\bf k}^\prime, p}(0) \right>
\nonumber
\\
&
= \left<\Gamma^{\dag}_{{\bf k},n}(\tau) \Gamma^{\dag}_{{\bf k}^\prime ,t}(0)\right> \left< \Gamma_{{\bf k},m}(\tau)\Gamma_{{\bf k}^\prime, p}(0) \right>
+ \left<\Gamma^{\dag}_{{\bf k},n}(\tau)\Gamma_{{\bf k}^\prime, p}(0) \right> \left<\Gamma_{{\bf k},m}(\tau) \Gamma^{\dag}_{{\bf k}^\prime ,t}(0)\right>,
\end{align}
where correlators with the same time, and hence which are disconnected, vanish. Explicit expressions of the resulting pair correlators are as follows,
\begin{align}
\left<\Gamma^{\dag}_{{\bf k},n}(\tau) \Gamma^{\dag}_{{\bf k}^\prime ,t}(0)\right> 
\nonumber
&
= \delta_{{\bf k},-{\bf k}^\prime} \delta_{n,t-N} \Theta(N-n)\Theta(t-N) g(\varepsilon_{{\bf k},n})e^{\varepsilon_{{\bf k},n}\tau}
\nonumber
\\
& - \delta_{{\bf k},-{\bf k}^\prime} \delta_{n,t+N} \Theta(n-N)\Theta(N-t) g(-\varepsilon_{-{\bf k},n})e^{-\varepsilon_{-{\bf k},n}\tau}
\nonumber
\\
&
= i\delta_{{\bf k},-{\bf k}^\prime} \left( \sigma_{2} \right)_{nt} g[ ( \sigma_{3}\varepsilon_{{\bf k}} )_{nn}] 
e^{\left( \sigma_{3}\varepsilon_{{\bf k}} \right)_{nn}\tau},
\end{align}
\begin{align}
\left<\Gamma_{{\bf k},m}(\tau) \Gamma_{{\bf k}^\prime ,p}(0)\right> 
&= - \delta_{{\bf k},-{\bf k}^{\prime}}\delta_{m,p-N}\Theta( N-m)\Theta(p-N) g(-\varepsilon_{{\bf k},m})e^{-\varepsilon_{{\bf k},m}\tau}
\nonumber
\\
&
+ \delta_{{\bf k},-{\bf k}^{\prime}}\delta_{p,m-N}\Theta( m-N)\Theta(N-p) g(\varepsilon_{-{\bf k},m})e^{\varepsilon_{-{\bf k},m}\tau}
\nonumber
\\
&
=- i\delta_{{\bf k},-{\bf k}^\prime} \left( \sigma_{2} \right)_{mp} g[-( \sigma_{3}\varepsilon_{{\bf k}} )_{mm}]
 e^{-\left( \sigma_{3}\varepsilon_{{\bf k}} \right)_{mm}\tau},
\end{align}
\begin{align}
\left<\Gamma^{\dag}_{{\bf k},n}(\tau) \Gamma_{{\bf k}^\prime ,p}(0)\right>
&
= \delta_{n,p} \delta_{{\bf k},{\bf k}^\prime} \Theta(N-n) \Theta(N-p) g(\varepsilon_{{\bf k},n}) e^{\varepsilon_{{\bf k},n}\tau } 
\nonumber
\\
&
- \delta_{n,p} \delta_{{\bf k},{\bf k}^\prime} \Theta(n-N)\Theta(p-N) g(-\varepsilon_{-{\bf k},n}) e^{-\varepsilon_{-{\bf k},n}\tau } 
\nonumber
\\
&
= \delta_{{\bf k},{\bf k}^\prime} \left( \sigma_{3}\right)_{np} g[( \sigma_{3} \varepsilon_{{\bf k}} )_{nn}] 
e^{\left( \sigma_{3} \varepsilon_{{\bf k}} \right)_{nn} \tau},
\end{align}
\begin{align}
\left<\Gamma_{{\bf k},m}(\tau)\Gamma^{\dag}_{{\bf k}^\prime ,t}(0) \right>
&
= - \delta_{m,t}\delta_{{\bf k},{\bf k}^\prime} \Theta(N-m)\Theta(N-t) g(-\varepsilon_{{\bf k},m}) e^{-\varepsilon_{{\bf k},m}\tau}
\nonumber
\\
&
+ \delta_{m,t}\delta_{{\bf k},{\bf k}^\prime} \Theta(m-N)\Theta(t-N) g(\varepsilon_{-{\bf k},m}) e^{\varepsilon_{-{\bf k},m}\tau}
\nonumber
\\
&
= - \delta_{{\bf k},{\bf k}^\prime} \left( \sigma_{3} \right)_{mt} g[ - \left(\sigma_{3}\varepsilon_{{\bf k}} \right)_{mm}] 
e^{-\left( \sigma_{3}\varepsilon_{{\bf k}}\right)_{mm}\tau},
\end{align}
where $g(\eta) = (e^{\eta/T} - 1)^{-1}$ is the Bose-Einstein distribution function. We then obtain
\begin{align}
S_{\alpha\beta} 
&= - \frac{1}{4} \lim_{\omega \to 0} \frac{\partial}{\partial \omega}\int_{0}^{1/T} d\tau e^{i\omega\tau} \sum_{{\bf k}{\bf k}^\prime} 
( {\tilde V}_{\alpha \bf k } )_{nm} \left(\varepsilon_{{\bf k}^\prime}\sigma_{3}{\tilde v}_{\beta {\bf k}^\prime}
+ {\tilde v}_{\beta {\bf k}^\prime}\sigma_{3}\varepsilon_{{\bf k}^\prime} \right)_{tp}
\nonumber
\\
&
\times \left[ 
\delta_{{\bf k},-{\bf k}^\prime} \left(\sigma_{2} \right)_{nt} \left(\sigma_{2} \right)_{mp} 
-
\delta_{{\bf k},{\bf k}^\prime} \left(\sigma_{3} \right)_{np}\left( \sigma_3 \right)_{mt}
\right] 
g[ ( \sigma_{3}\varepsilon_{{\bf k}})_{nn}] g[ -( \sigma_{3}\varepsilon_{{\bf k}})_{mm} ]
e^{\left(\sigma_{3}\varepsilon_{{\bf k}} \right)_{nn}\tau } e^{-\left(\sigma_{3}\varepsilon_{{\bf k}} \right)_{mm}\tau}.
\end{align}
Integrate over time (recall that $\omega$ is a bosonic Matsubara frequency)
\begin{align}
\int_{0}^{1/T} e^{i\omega \tau} e^{\left(\sigma_{3}\varepsilon_{{\bf k}} \right)_{nn}\tau} e^{-\left(\sigma_{3}\varepsilon_{{\bf k}} \right)_{mm}\tau}
= \frac{e^{\left(\sigma_{3}\varepsilon_{{\bf k}} \right)_{nn}/T} e^{-\left(\sigma_{3}\varepsilon_{{\bf k}} \right)_{mm}/T} - 1}
{i\omega + \left(\sigma_{3}\varepsilon_{{\bf k}} \right)_{nn}-\left(\sigma_{3}\varepsilon_{{\bf k}} \right)_{mm}}.
\end{align}
An identity 
\begin{align}
-\left[ e^{\left(\sigma_{3}\varepsilon_{{\bf k}} \right)_{nn}/T} e^{-\left(\sigma_{3}\varepsilon_{{\bf k}} \right)_{mm}/T} - 1 \right]
g[\left(\sigma_{3}\varepsilon_{{\bf k}} \right)_{nn} ] g[ - \left(\sigma_{3}\varepsilon_{{\bf k}} \right)_{mm} ]
= g[\left(\sigma_{3}\varepsilon_{{\bf k}} \right)_{nn} ] - g[ \left(\sigma_{3}\varepsilon_{{\bf k}} \right)_{mm} ]
\end{align}
is certainly useful. We then get
\begin{align}
S_{\alpha\beta} = 
&
\frac{1}{4} \lim_{\omega \to 0} \frac{\partial}{\partial \omega}\sum_{{\bf k}{\bf k}^\prime} 
( {\tilde V}_{\alpha \bf k} )_{nm} \left( \varepsilon_{{\bf k}^\prime}\sigma_{3}{\tilde v}_{\beta {\bf k}^\prime}
+ {\tilde v}_{\beta {\bf k}^\prime}\sigma_{3}\varepsilon_{{\bf k}^\prime} \right)_{tp}
\nonumber
\\
&
\left[ 
\delta_{{\bf k},-{\bf k}^\prime} \left(\sigma_{2} \right)_{nt}\left(\sigma_{2} \right)_{mp} 
-
\delta_{{\bf k},{\bf k}^\prime} \left(\sigma_{3} \right)_{np}\left( \sigma_3 \right)_{mt}
\right] 
\frac{g[\left(\sigma_{3}\varepsilon_{{\bf k}} \right)_{nn} ] - g[ \left(\sigma_{3}\varepsilon_{{\bf k}} \right)_{mm} ]}
{i\omega + \left(\sigma_{3}\varepsilon_{{\bf k}} \right)_{nn}-\left(\sigma_{3}\varepsilon_{{\bf k}} \right)_{mm}}.
\end{align}

Let us focus on the $\delta_{{\bf k},-{\bf k}^\prime}\sigma_{2}\sigma_{2}$ of the Kubo formula, and show it doubles the $\delta_{{\bf k},{\bf k}^\prime}\sigma_{3}\sigma_{3}$ term.
We use $T_{{\bf k}}^{\dag}= P_{\bf k}^{\dag} (\sigma_{1}T_{-{\bf k}}^{\mathrm{T}} \sigma_{1})$ and $H_{\bf k} = \sigma_{1}H_{-\bf k}^{\mathrm{T}}\sigma_{1}$ relations to show
\begin{align}
{\tilde v}_{\beta,-{\bf k}}& = - T_{-{\bf k}}^{\dag}\left( \partial_{\beta}H_{-{\bf k}} \right) T_{-{\bf k}}
= - P_{- \bf k}^{\dag}\left[ \sigma_{1}T_{\bf k}^{\dag}\sigma_{1}\left( \partial_{\beta}H_{-\bf k}^{\mathrm{T}}\right)\sigma_{1}T_{\bf k}\sigma_{1} \right]^{\mathrm{T}}
 P_{-\bf k}
\nonumber
\\
&
=  - P_{- \bf k}^{\dag}\sigma_{1} \left[ T_{\bf k}^{\dag} \left( \partial_{\beta}H_{\bf k}\right) T_{\bf k} \right]^{\mathrm{T}} \sigma_{1}P_{-\bf k}
=  - P_{- \bf k}^{\dag}\sigma_{1} {\tilde v}_{\beta {\bf k}}^{\mathrm{T}} \sigma_{1}P_{-\bf k}.
\end{align}
With that and $\varepsilon_{-{\bf k}} = \sigma_{1}\varepsilon_{\bf k}\sigma_{1}$ we show
\begin{align}
\left( \varepsilon_{-\bf k}\sigma_{3}{\tilde v}_{\beta, - {\bf k}}
+ {\tilde v}_{\beta, - {\bf k}}\sigma_{3}\varepsilon_{- \bf k} \right)_{tp}
\left( \sigma_{2} \right)_{nt}\left(\sigma_{2} \right)_{mp}  = 
-\left( \sigma_{3}P_{\bf k}{\tilde v}_{\beta {\bf k}}P_{\bf k}^{\dag}\varepsilon_{\bf k} + \varepsilon_{\bf k}P_{\bf k}{\tilde v}_{\beta {\bf k}}P_{\bf k}^{\dag}\sigma_{3} \right)_{mn},
\end{align}
where we used $(\Lambda^{\mathrm{T}})_{nm} = \Lambda_{mn}$, and $\sigma_{1}P_{-{\bf k}}^{*}\sigma_{1} = P_{\bf k}^{\dag}$ identities. By redefining the $T_{\bf k}$ matrix as 
\begin{align}
{\tilde T}_{\bf k} = T_{\bf k}P_{\bf k}^{\dag}
\end{align}
we get
\begin{align}
-\left( \sigma_{3}P_{\bf k}{\tilde v}_{\beta {\bf k}}P_{\bf k}^{\dag}\varepsilon_{\bf k} + \varepsilon_{\bf k}P_{\bf k}{\tilde v}_{\beta {\bf k}}P_{\bf k}^{\dag}\sigma_{3} \right)_{mn}  = - \left( \sigma_{3}{\bar v}_{\beta \bf k}\varepsilon_{\bf k} + \varepsilon_{\bf k}{\bar v}_{\beta \bf k}\sigma_{3} \right)_{mn},
\end{align}
where now
\begin{align}
{\bar v}_{\beta \bf k} = {\tilde T}^{\dag}_{\bf k}\left( \partial_{\beta}H_{\bf k} \right) {\tilde T}_{\bf k}
\end{align}
We could have from the very beginning chosen such a $T_{\bf k}$ in the $\delta_{{\bf k},-{\bf k}^\prime}\sigma_{2}\sigma_{2}$ term of the Kubo formula that all $P_{\bf k}$ matrices get absorbed.
Hence the $\delta_{{\bf k},-{\bf k}^\prime}\sigma_{2}\sigma_{2}$ term of the Kubo formula doubles the  $\delta_{{\bf k},{\bf k}^\prime}\sigma_{3}\sigma_{3}$ term.

Taking a derivative with respect to $\omega$ and then setting $\omega = 0$, we get
\begin{align}
S_{\alpha\beta} = 
&
 \frac{i}{2} \sum_{{\bf k} n} 
( {\tilde V}_{\alpha{\bf k}} )_{nm} \left(  \varepsilon_{\bf k}\sigma_{3}{\tilde v}_{\beta{\bf k}}
+ {\tilde v}_{\beta{\bf k}}\sigma_{3}\varepsilon_{\bf k} \right)_{mn}
 \left(\sigma_{3} \right)_{nn}\left( \sigma_3 \right)_{mm}
\frac{g[\left(\sigma_{3}\varepsilon_{\bf k} \right)_{nn} ] - g[ \left(\sigma_{3}\varepsilon_{{\bf k}} \right)_{mm} ]}
{\left[ \left(\sigma_{3}\varepsilon_{{\bf k}} \right)_{nn}-\left(\sigma_{3}\varepsilon_{{\bf k}} \right)_{mm}\right]^2}
\end{align}

Finally, after all transformation, one obtains for the transverse part of the Kubo formula
\begin{align}
S^{[\mathrm{B}]}_{\alpha\beta} = \frac{i}{2} \sum_{{\bf k}n} \int_{-\infty}^{+\infty} d\eta g(\eta)
\left[ {\bar O} \partial_{\alpha}T_{\bf k}^{\dag} \sigma_{3}\left( \eta + \sigma_{3}H_{\bf k} \right)\partial_{\beta}T_{\bf k} 
 \right]_{nn}  \delta\left[ \eta - \left( \sigma_{3}\varepsilon_{\bf k}\right)_{nn} \right]  - \left( \alpha \leftrightarrow \beta \right),
\end{align}
here ${\bar O} = \sigma_{3} T_{\bf k}^{\dag}{\hat O}T_{\bf k}\sigma_{3}$, and index $\mathrm{B}$ stands for the Berry curvature contribution, i.e. transverse part of the Kubo formula.

\subsection{Magnon orbital magnetization part}
In Fourier space, the magnetization part of the current is given by 
\begin{align}
 J_{\alpha}^{[1]}  = \frac{1}{2V}\mathrm{Tr} \sum_{{\bf k}} \sigma_{3}{\hat O}\sigma_{3} \left( r_{\beta}v_{\alpha {\bf k}} + v_{\alpha{\bf k}}r_{\beta} \right)g(\sigma_{3}\varepsilon_{\bf k}) \nabla_{\beta}\chi \equiv \frac{1}{V} M_{\alpha\beta}\nabla_{\beta}\chi .
\end{align}
Our goal now is to derive an expression for $M_{\alpha\beta}$.
For that, following Smrcka and Streda Ref. \onlinecite{Smrcka.Streda:JPC1977} and adopting calculations from Ref. \onlinecite{Matsumoto.Shindou.ea:PRB2014}, we introduce two helpful functions
\begin{align}
&
A_{\alpha\beta}(\eta) = i \mathrm{Tr} 
\left[ \sigma_{3} V_{\alpha {\bf k}} \frac{d G^{+}}{d\eta} \sigma_{3} v_{\beta {\bf k}} \delta(\eta - \sigma_{3}H_{{\bf k}}) 
- \sigma_{3} V_{\alpha {\bf k}} \delta(\eta - \sigma_{3}H_{{\bf k}}) \sigma_{3} v_{\beta {\bf k}} \frac{d G^{-}}{d\eta} \right],
\\
&
B_{\alpha\beta}(\eta) = i\mathrm{Tr} 
\left[ \sigma_{3} V_{\alpha {\bf k}}G^{+}\sigma_{3}v_{\beta {\bf k}}\delta(\eta - \sigma_{3}H_{{\bf k}}) 
- \sigma_{3} V_{\alpha {\bf k}}\delta(\eta - \sigma_{3}H_{{\bf k}}) \sigma_{3}v_{\beta {\bf k}}G^{-} \right],
\end{align}
where $G^{\pm} = \left( \eta^{\pm} - \sigma_{3} H_{\bf k}\right)^{-1}$ is the Green function where $\eta^{\pm} = \eta \pm 0$, and where we defined $V_{\alpha {\bf k}} = {\hat O}\sigma_{3}v_{\alpha {\bf k}}$. 
It is straightforward to show
\begin{align}
A_{\alpha\beta}(\eta) - \frac{1}{2}\frac{d B_{\alpha\beta} (\eta)}{d\eta}
&
= \frac{1}{2 \pi}\mathrm{Tr} \left\{ \sigma_{3}{\hat O} \left[ x_{\alpha}x_{\beta} \left( G^{+} - G^{-} \right) - x_{\alpha} \left( G^{+} - G^{-} \right)x_{\beta} \right]  \right\}
\nonumber
\\
&
- \frac{i}{4\pi} \mathrm{Tr} 
\left\{ \sigma_{3}V_{\alpha {\bf k} } \left[ (G^{+})^2 - (G^{-})^2 \right] x_{\beta}
 + \sigma_{3}V_{\alpha {\bf k} } x_{\beta} \left[ (G^{+})^2 - (G^{-})^2 \right]  \right\},
\end{align}
where we used $v_{\alpha {\bf k} } = i[x_{\alpha},\sigma_{3}\left( G^{\pm}\right)^{-1}]$ and ${\hat O}\sigma_{3}{\hat H} = {\hat H}\sigma_{3}{\hat O}$ assumption.
Identities
\begin{align}
&
G^{+} - G^{-} = -2\pi i \delta\left( \eta - \sigma_{3}H_{{\bf k}}\right),
\\
&
(G^{+})^2 - (G^{-})^2 = 2\pi i \frac{d}{d\eta}  \delta\left( \eta - \sigma_{3}H_{{\bf k}}\right),
\\
&
G^{\pm} = T_{\bf k}\sigma_{3} \frac{1}{\eta^{\pm} - \sigma_{3}\varepsilon_{\bf k}} T_{\bf k}^{\dag} \sigma_{3} ,
\end{align}
are then used to obtain
\begin{align}
A_{\alpha\beta}(\eta) - \frac{1}{2}\frac{d B_{\alpha\beta} (\eta)}{d\eta} 
&= 
 -2i \mathrm{Tr} \left\{ \sigma_{3}{\hat O} \left[ x_{\alpha}x_{\beta} \delta\left( \eta - \sigma_{3}H_{{\bf k}}\right) 
- x_{\alpha} \delta\left( \eta - \sigma_{3}H_{{\bf k}}\right) x_{\beta} \right]  \right\}
\nonumber
\\
&+ 
\frac{1}{2} \mathrm{Tr} 
\left[ \sigma_{3}V_{\alpha {\bf k} }  \frac{d}{d\eta}  \delta\left( \eta - \sigma_{3}H_{{\bf k}}\right)  x_{\beta}
 + \sigma_{3}V_{\alpha {\bf k} } x_{\beta}  \frac{d}{d\eta}  \delta\left( \eta - \sigma_{3}H_{{\bf k}}\right)  \right].
\end{align}
The first line vanishes due to $[x_{\alpha},x_{\beta}] = 0$. We can deduce that operator 
${\bar O} = \sigma_{3} T_{\bf k}^{\dag}{\hat  O}T_{\bf k}\sigma_{3}$ is diagonal. That can be seen from the commutation relation
${\bar O} \varepsilon_{\bf k}\sigma_{3} = \sigma_{3}\varepsilon_{\bf k}{\bar  O} $, which in the diagonal basis are rewritten as
${\bar O}_{nm} \left[  \left( \varepsilon_{\bf k}\sigma_{3} \right)_{mm} - \left(\varepsilon_{\bf k}\sigma_{3} \right)_{nn}   \right] = 0$. 
We can deduce another two useful identities,
\begin{align}
&
{\tilde v}_{\alpha {\bf k}} {\bar O} = \sigma_{3} {\bar O} {\tilde v}_{\alpha {\bf k}} \sigma_{3},
\\
&
{\bar O} \partial_{\alpha}T_{\bf k}^{\dag} = \sigma_{3} \partial_{\alpha}T_{\bf k}^{\dag}{\hat O}\sigma_{3}.
\end{align}
An expression for the velocity written in the diagonal basis 
\begin{align}
\left( {\tilde v}_{\alpha {\bf k}}\right)_{nm} = \left( \partial_{\alpha} \varepsilon_{\bf k} \right)_{nm} 
+ \left( \mathcal{A}_{\alpha{\bf k}} \right)_{nm} \left[  \left(\sigma_{3}\varepsilon_{\bf k} \right)_{mm} - \left( \sigma_{3}\varepsilon_{\bf k} \right)_{nn}\right],
\end{align}
where $\mathcal{A}_{\alpha{\bf k}} = T_{\bf k}^{\dag}\sigma_{3}\partial_{\alpha}T_{\bf k}$, will be used in the following derivations.
It is necessary to explicitly write down an expression for the Berry curvature part of $A_{\alpha\beta}$ as
\begin{align}
A^{[\mathrm{B}]}_{\alpha\beta} (\eta)
= - i \sum_{n}   \left( {\bar O} \partial_{\alpha}T_{\bf k}^{\dag} \sigma_{3} \partial_{\beta}T_{\bf k} \right)_{nn} 
\delta\left[ \eta  - \left( \sigma_{3}\varepsilon_{\bf k}\right)_{nn}\right] - (\alpha \leftrightarrow \beta),
\end{align}
and a Berry curvature part of the $B_{\alpha\beta}$ as
\begin{align}
B^{[\mathrm{B}]}_{\alpha\beta} (\eta) 
= i \sum_{n}
\left[ {\bar O} \partial_{\alpha}T_{\bf k}^{\dag}\sigma_{3}\left(\eta - \sigma_{3}H_{\bf k} \right) \partial_{\beta}T_{\bf k} \right]_{nn} 
 \delta\left[ \eta - \left( \sigma_{3}\varepsilon_{\bf k}\right)_{nn}\right] - \left( \alpha \leftrightarrow \beta\right).
\end{align}
For bounded spectrum, we derive a sum rule
\begin{align}
\int_{-\infty}^{\infty} d\eta \left[ A_{\alpha\beta}(\eta) - \frac{1}{2}\frac{d B_{\alpha\beta} (\eta)}{d\eta} \right]
= \sum_{n}\left( {\bar O}\partial_{\alpha}T_{\bf k}^{\dag}\sigma_{3}\partial_{\beta}T_{\bf k}\right)_{nn} - (\alpha \leftrightarrow \beta) = 0.
\end{align} 
It can then be shown that the $M_{\alpha\beta}$ is expressed through $A_{\alpha\beta}$ and $B_{\alpha\beta}$ as
\begin{align}
&
M_{\alpha\beta} = \sum_{{\bf k}} \int_{-\infty}^{\infty} d{\tilde \eta} \left[ A_{\alpha\beta}({\tilde \eta}) - \frac{1}{2} \frac{d B_{\alpha\beta}({\tilde \eta})}{d {\tilde \eta}}\right]
\int_{0}^{{\tilde \eta}}d\eta g(\eta)
\nonumber
\\
&
= -i\sum_{{\bf k}n}\int_{-\infty}^{\infty} d{\tilde \eta} \left( {\bar O}\partial_{\alpha}T_{\bf k}^{\dag} \sigma_{3} \partial_{\beta}T_{\bf k} \right)_{nn} 
\delta\left[ {\tilde \eta} - \left(\sigma_{3}\varepsilon_{\bf k}\right)_{nn} \right] \int_{0}^{{\tilde \eta}} d\eta g(\eta)
\nonumber
\\
&
+ \frac{i}{2} \sum_{{\bf k}n}\int_{-\infty}^{\infty} d{\tilde \eta} \left[ {\bar O}\partial_{\alpha}T_{\bf k}^{\dag} \sigma_{3}({\tilde \eta} - \sigma_{3}H_{\bf k}) \partial_{\beta}T_{\bf k} \right]_{nn} g({\tilde \eta}) \delta\left[ {\tilde \eta} - \left(\sigma_{3}\varepsilon_{\bf k}\right)_{nn} \right]
\end{align}

\subsection{Overall response}
Overall response of the current on the temperature gradient is summarized as $J_{\alpha} = \frac{1}{V} L_{\alpha\beta}\nabla_{\beta}\chi$, where
\begin{align}
L_{\alpha\beta} = S_{\alpha\beta} + M_{\alpha\beta} = \sum_{{\bf k}n} \left[ {\bar \Omega}^{[\mathrm{O}]}_{\alpha\beta}({\bf k})\right]_{nn}
\int_{0}^{\left(\sigma_{3}\varepsilon_{\bf k}\right)_{nn}} d\eta \eta \frac{dg(\eta)}{d\eta} ,
\end{align}
where 
\begin{align}
{\bar \Omega}^{[\mathrm{O}]}_{\alpha\beta}({\bf k}) = i  {\bar O}\partial_{\alpha}T_{\bf k}^{\dag}\sigma_{3}\partial_{\beta}T_{\bf k}   
- \left( \alpha \leftrightarrow \beta\right)
\end{align}
is the Berry curvature with an operator ${\bar O}$, dub it $O-$Berry curvature. As was shown above, the Berry curvature must satisfy $\mathrm{Tr}{\bar \Omega}^{[\mathrm{O}]}_{\alpha\beta}({\bf k}) =0 $ sum rule.

\section{A model}
Here we introduce a model of an antiferromagnet on honeycomb lattice.
Assume that the Neel order is in $z-$ direction, and allow second nearest neighbor Dzyaloshinskii-Moriya interaction. 
The Hamiltonian is
\begin{align}
H = J\sum_{<ij>}{\bf S}_{i}{\bf S}_{j} +  D \sum_{<<ij>>} \nu_{ij}\left[ {\bf S}_{i}\times {\bf S}_{j}\right]_{z}.
\end{align} 
Upon Holstein-Primakoff transformation, performing Fourier transformation, we arrive at a Hamiltonian
\begin{align}
H_{\bf k} =  JS \left[
\begin{array}{cccc}
3 +\Delta_{\bf k} & 0 & 0 & - \gamma_{\bf k} \\
0 & 3 + \Delta_{\bf k} & - \gamma_{- \bf k} & 0 \\
0 & - \gamma_{\bf k} & 3 -\Delta_{\bf k} & 0 \\
- \gamma_{-\bf k} & 0 & 0 & 3 - \Delta_{\bf k}
\end{array}
\right],
\end{align}
with a spinor $\Psi_{\bf k} = (a_{\bf k},b_{\bf k}, a^{\dag}_{-\bf k}, b^{\dag}_{-\bf k})^{\mathrm{T}}$. We defined $\gamma_{\bf k} = \sum_{i}e^{i{\bf k}{\bm \tau}_{i}}$ where ${\bm \tau}_{1} = \frac{1}{2}(\frac{1}{\sqrt{3}},1 )$,
 ${\bm \tau}_{2} = \frac{1}{2}(\frac{1}{\sqrt{3}},-1)$, and ${\bm \tau}_{3} = \frac{1}{\sqrt{3}}(-1,0 )$, hence
\begin{align}
\gamma_{\bf k}  =  2e^{ik_{x}\frac{1}{2\sqrt{3}}}\cos\left( \frac{k_{y}}{2}\right) + e^{-ik_{x}\frac{1}{\sqrt{3}}}.
\end{align}
We then defined $\Delta_{\bf k} = 2\Delta \left[ -\sin({\bf k}{\bf a}_{1}) + \sin({\bf k}{\bf a}_{2}) + \sin({\bf k}{\bf a}_{1} - {\bf k}{\bf a}_{2}) \right]$, where $\Delta = D/J$, and
${\bf a}_{1} = \frac{1}{2}(\sqrt{3},1)$, and ${\bf a}_{2} = \frac{1}{2}(\sqrt{3}, -1)$, and we get
\begin{align}
\Delta_{{\bf k}} = 2\Delta \left[\sin(k_{y}) - 2\sin\left( \frac{k_{y}}{2}\right)\cos\left(\frac{\sqrt{3}k_{x}}{2} \right) \right],
\end{align}
hence $\Delta_{{\bf k}} = - \Delta_{-{\bf k}}$.
We define spin density operator 
\begin{align}
{\hat O} = \left[
\begin{array}{cc}
{\hat \tau}_{3} & 0 \\
0 & {\hat \tau}_{3} 
\end{array} \right],
\end{align} 
where ${\hat \tau}_{3}$ is third $2\times 2$ Pauli matrix. 
It can be easily checked that this operator satisfies ${\hat H}\sigma_{3}{\hat O} - {\hat O}\sigma_{3}{\hat H} = 0$ condition for existence of a well defined current. 

\begin{figure} \centerline{\includegraphics[clip,width=0.35\columnwidth]{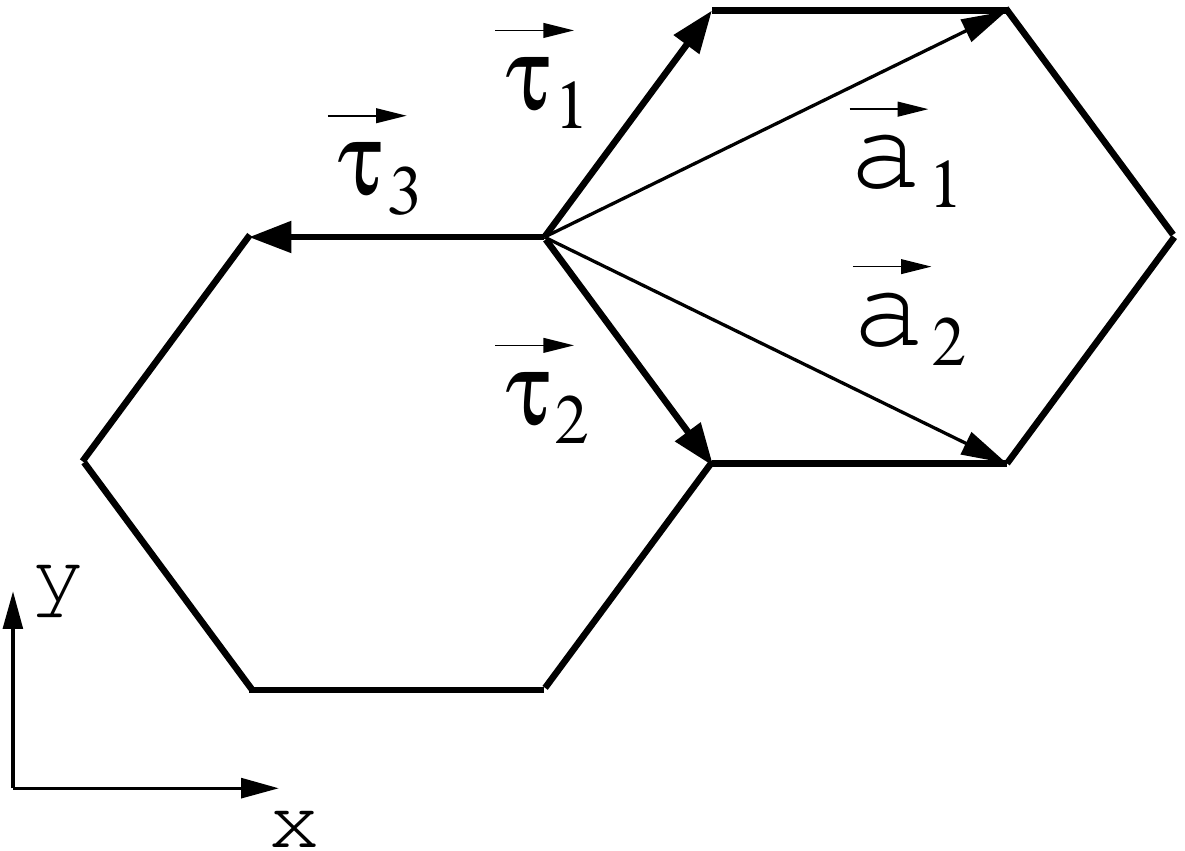}}

\protect\caption{Schematics of the graphene layer parametres for the tight-binding model. }

\label{fig:Fig1}  

\end{figure}

For the block $\mathrm{\Rmnum{1}}$ described by
\begin{align}
\Psi_{\mathrm{\Rmnum{1}}} = \left[ 
\begin{array}{c}
a_{\bf k} \\
b^{\dag}_{-{\bf k}}
\end{array}
\right]
\end{align}
spinor the Hamilonian is
\begin{align}
H_{ \mathrm{\Rmnum{1}} \bf k} = 
JS \left[ \begin{array}{cc}
3 +\Delta_{\bf k} & -\gamma_{\bf k} \\
-\gamma_{-\bf k} & 3 - \Delta_{\bf k}
\end{array} \right]
\end{align}
Upon diagonalization of the Hamiltonian, we get the spectrum
\begin{align}
E_{ \bf k }  = JS\left(  \Delta_{\bf k}  +  \sqrt{ 9 - \vert \gamma_{\bf k} \vert^2  } \right)
\end{align}
with corresponding eigenfunctions
\begin{align}
\Psi_{ \mathrm{\Rmnum{1}} +} =   \left[ \begin{array}{c}
\cosh(\xi_{  {\bf k} }/2 ) e^{i\chi_{{\bf k}}} \\
\sinh(\xi_{  {\bf k}}/2 ) \end{array} \right],
~~
\Psi_{ \mathrm{\Rmnum{1}} -} =   \left[ \begin{array}{c}
\sinh(\xi_{  {\bf k} }/2 )  \\
\cosh(\xi_{  {\bf k}}/2 )e^{-i\chi_{{\bf k}}} \end{array} \right],
\end{align}
where $\gamma_{\bf k} = \vert \gamma_{\bf k}\vert e^{i\chi_{\bf k}}$, and 
\begin{align}
\sinh(\xi_{ \bf k}) = \frac{\vert \gamma_{\bf k} \vert}{\epsilon_{ \bf k}}, 
~~ \cosh(\xi_{ \bf k}) = \frac{3}{\epsilon_{  \bf k}},
\end{align}
in which $\epsilon_{ \bf k} =   \sqrt{ 9 - \vert \gamma_{\bf k} \vert^2  }$ was defined. To be specific, the eigenvector $\Psi_{ \mathrm{\Rmnum{1}} +} $ corresponds to $E_{\bf k}$ eigenvalue, while $\Psi_{ \mathrm{\Rmnum{1}} -} $ to $E_{-\bf k}$.
Matrix $T_{ \mathrm{\Rmnum{1}} {\bf k}}$ is readily constructed to be
\begin{align}
T_{ \mathrm{\Rmnum{1}} {\bf k}} = \left[
\begin{array}{cc}
\cosh(\xi_{  {\bf k} }/2 ) e^{i\chi_{{\bf k}}} & \sinh(\xi_{  {\bf k}}/2 )   \\
\sinh(\xi_{ {\bf k}}/2 )  & \cosh(\xi_{  {\bf k} }/2 ) e^{-i\chi_{{\bf k}}}
\end{array} \right].
\end{align}
It can be checked that indeed
\begin{align}
T_{\bf k}^{\dag} H_{\bf k} T_{\bf k} = 
\left[ 
\begin{array}{cc}
E_{\bf k} & 0 \\
0 & E_{-\bf k} 
\end{array}\right],
\end{align}
as defined above.

The $O$-Berry curvature for the $\mathrm{\Rmnum{1}}$ block of the Hamiltonian is 
\begin{align}
\left({\bar \Omega}_{\alpha\beta}^{[\mathrm{O}]} \right)_{\mathrm{\Rmnum{1}}} 
=i{\bar O}_{\mathrm{\Rmnum{1}}} \left(\partial_{\alpha}T_{ \mathrm{\Rmnum{1}} {\bf k}}^{\dag}\right) \tau_{3} \left(\partial_{\beta} T_{ \mathrm{\Rmnum{1}} {\bf k}} \right) 
- (\alpha \leftrightarrow\beta).
\end{align}
It can be shown that 
${\bar O}_{\mathrm{\Rmnum{1}}} = \tau_{3} T_{ \mathrm{\Rmnum{1}} {\bf k}}^{\dag} {\hat O}_{\mathrm{\Rmnum{1}}} T_{ \mathrm{\Rmnum{1}} {\bf k}} \tau_{3} = \tau_{3}$, hence
\begin{align}
\left( {\bar \Omega}_{\alpha\beta}^{[\mathrm{O}]} \right)_{\mathrm{\Rmnum{1}}} =  
i \tau_{3}  \left(\partial_{\alpha}T_{ \mathrm{\Rmnum{1}} {\bf k}}^{\dag}\right) \tau_{3} \left(\partial_{\beta} T_{ \mathrm{\Rmnum{1}} {\bf k}} \right) 
- (\alpha \leftrightarrow\beta)  .
\end{align}
For the sake of calculating the current, we will be needing only the diagonal parts of the curvature. 
The two diagonal elements are expressed by
\begin{align}
 \left\{ \left( {\bar \Omega}_{\alpha\beta}^{[\mathrm{O}]} \right)_{\mathrm{\Rmnum{1}}} \right\}_{11}=
 -\left\{ \left( {\bar \Omega}_{\alpha\beta}^{[\mathrm{O}]} \right)_{\mathrm{\Rmnum{1}}} \right\}_{22} 
  \equiv
  \Omega_{\alpha\beta}^{[\mathrm{O}]} 
= - \frac{1}{2} \sinh(\xi_{  \bf k}) \left[ \partial_{\beta}\chi_{\bf k} \partial_{\alpha} \xi_{  \bf k}
- \partial_{\alpha}\chi_{\bf k}\partial_{\beta}\xi_{ \bf k}  \right].
\end{align}
Using following identities
\begin{align}
&
\partial_{\alpha} \chi_{\bf k} = \frac{1}{\vert \gamma_{\bf k} \vert^2} 
\left( \mathrm{Re}\gamma_{\bf k}\partial_{\alpha}\mathrm{Im}\gamma_{\bf k} 
-
\mathrm{Im}\gamma_{\bf k}\partial_{\alpha}\mathrm{Re}\gamma_{\bf k}
 \right),
\\
&
\partial_{\alpha}\xi_{ \bf k} = \frac{3}{\epsilon^2_{ \bf k}}
 \partial_{\alpha}\vert \gamma_{\bf k}\vert ,
\end{align}
we show 
\begin{align}
& 
\partial_{\beta}\chi_{\bf k}\partial_{\alpha}\xi_{\bf k} - \partial_{\alpha}\chi_{\bf k}\partial_{\beta}\xi_{\bf k} 
\\
&
 = 
\frac{3}{\epsilon_{  \bf k}^{2}\vert\gamma_{\bf k}\vert}
\left[ \left( \partial_{\alpha}\mathrm{Re}\gamma_{\bf k}\right)\left( \partial_{\beta}\mathrm{Im}\gamma_{\bf k}\right) 
- \left(\partial_{\beta}\mathrm{Re}\gamma_{\bf k} \right)\left(\partial_{\alpha}\mathrm{Im}\gamma_{\bf k} \right)   \right],
\end{align}
such that final general expression for the $\mathrm{O}-$Berry curvature diagonal elements is 
\begin{align}
\Omega_{\alpha\beta}^{[\mathrm{O}]}
= -  \frac{3}{2 \epsilon_{  \bf k}^{3}}
\left[ \left( \partial_{\alpha}\mathrm{Re}\gamma_{\bf k}\right)\left( \partial_{\beta}\mathrm{Im}\gamma_{\bf k}\right) 
- \left(\partial_{\beta}\mathrm{Re}\gamma_{\bf k} \right)\left(\partial_{\alpha}\mathrm{Im}\gamma_{\bf k} \right)   \right].
\end{align}
Second, $\mathrm{\Rmnum{2}}$, block of the Hamiltonian can be obtained by a straightforward replacement of $\chi_{\bf k} \rightarrow - \chi_{\bf k}$ in the results above and recalling that ${\bar O}_{\mathrm{\Rmnum{2}}} = -\tau_{3}$. That does not change the Berry curvature for the $\mathrm{\Rmnum{2}}$ block comparing to the one obtained for the $\mathrm{\Rmnum{1}}$ block. The Berry curvature is plotted in Fig. [\ref{fig:Fig2}].

We are now in position to derive the spin current. A general expression is given by
\begin{align}
\left[ {\bf J}_{\mathrm{O}} \right]_{\alpha} =- \frac{2}{V}\sum_{\bf k}  \Omega^{[\mathrm{O}]}_{\alpha\beta}({\bf k})\left[ c_{1}(E_{\bf k}) - c_{1}(E_{-\bf k})\right] \nabla_{\beta}\chi ,
\end{align}
where $c_{1}(x) = \int_{0}^{x} d\eta~\eta\frac{dg(\eta)}{d\eta}$, where $g(\eta) =\left(e^{\beta\eta} - 1 \right)^{-1}$ is the Bose-Einstein distribution function with $\beta = 1/T$, and $V$ is the volume of the system. It is the asymmetry between the $E_{\bf k}$ and $E_{-\bf k}$ that results in non-zero spin Nernst current.
We would like to extract some analytic results.

\begin{figure}[h] \includegraphics[width=0.6\linewidth]{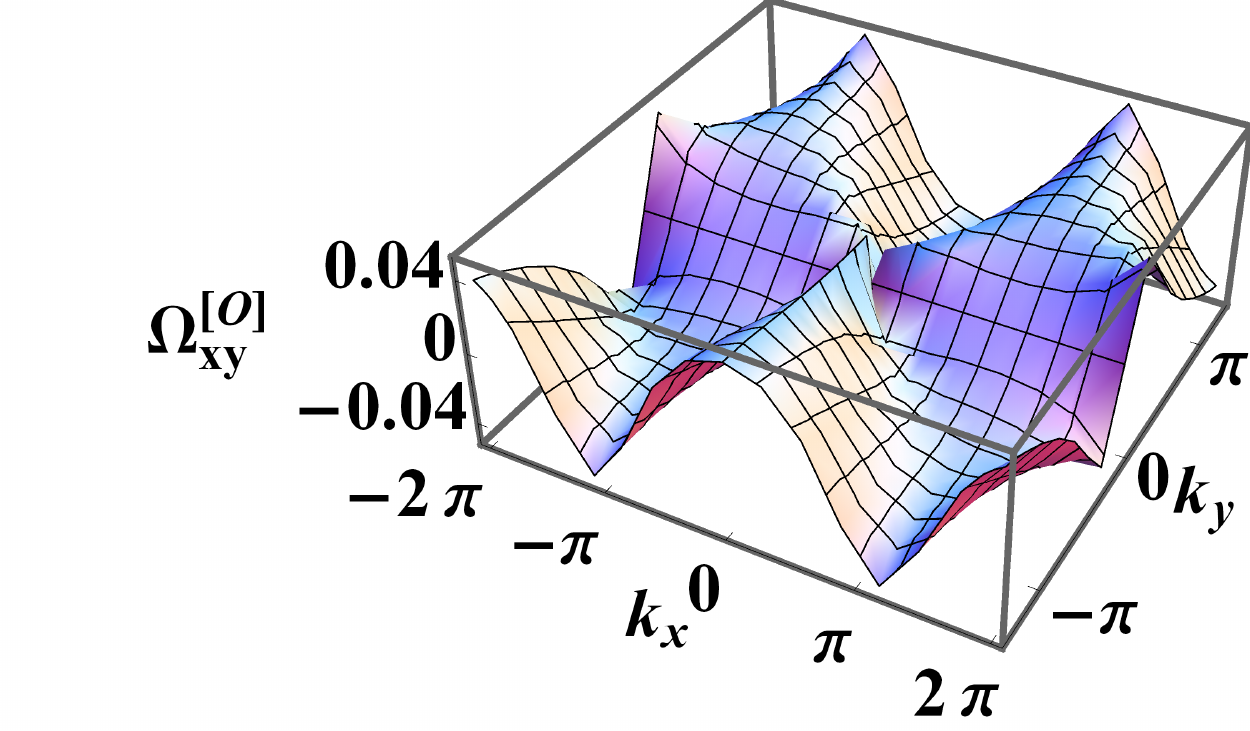}

\protect\caption{(Color online)Berry curvature of a single layer honeycomb antiferromagnet. The DMI does not affect the Berry curvature.}

\label{fig:Fig2}  

\end{figure}
\subsection{$\Gamma$ point}
We note that since the $\Gamma = (0,0)$ point  is not gapped, it will contribute the most to the spin current at low temperatures. 
We again consider only the ${\mathrm{\Rmnum{1}} }$ block.
We then expand all functions entering the current close to $\Gamma$ point in small ${\bf k}$ as
\begin{align}
&
\Delta_{\bf k} \approx \frac{1}{4}\Delta k_{y}\left( 3k_{x}^2 - k_{y}^2\right)
\\
&
\mathrm{Re}\gamma_{\bf k} \approx 3 - \frac{1}{4}k^2,
\\
&
\mathrm{Im}\gamma_{\bf k} \approx \frac{1}{24\sqrt{3}}k_{x}\left( k_{x}^2 - 3k_{y}^2 \right),
\\
&
\epsilon_{\bf k} \approx \sqrt{\frac{3}{2}} k.
\end{align}
Using all these expansions, with
\begin{align}
\partial_{x}\mathrm{Re}\gamma_{\bf k}\partial_{y}\mathrm{Im}\gamma_{\bf k} - \partial_{y}\mathrm{Re}\gamma_{\bf k}\partial_{x}\mathrm{Im}\gamma_{\bf k} 
= \frac{\sqrt{3}}{48}k_{y}( 3k_{x}^2 - k_{y}^2 )
\end{align}
we can write down an expression for the Berry curvature diagonal elements
\begin{align}
  \Omega_{xy}^{[\mathrm{O}]}   = - \frac{\sqrt{2}}{48  k^3} k_{y}(3k_{x}^2 - k_{y}^2).
\end{align}
One can check that the second, $\mathrm{\Rmnum{2}}$, block of the Hamiltonian doubles the results of the $\mathrm{\Rmnum{1}}$ block studied.
Note that the Berry curvature does not depend on the Dzyaloshinskii-Moriya strength and is a property of a honeycomb lattice. Integral of the Berry curvature over the Brillouin zone vanishes. 
The current is then
\begin{align}
J_{x} = - \frac{2}{V} \sum_{{\bf k}}  \Omega^{[\mathrm{O}]}_{xy}({\bf k}) \left[ c_{1}(E_{\bf k}) - c_{1}(E_{-\bf k})  \right]\nabla_{y}\chi .
\end{align}
Assuming small DMI, $D<J$, we approximate 
\begin{align}
 c_{1}(E_{\bf k}) - c_{1}(E_{-\bf k})  = \int_{E_{-\bf k}}^{E_{ \bf k}}d\eta~\eta\frac{dg(\eta)}{d\eta}
\approx 2(JS)\Delta_{\bf k}\epsilon_{\bf k} \frac{dg(JS\epsilon_{\bf k})}{d\epsilon_{\bf k}},
\end{align}
and with a help of 
\begin{align}
\int_{0}^{\infty} \frac{z^{x-1}}{e^{z} - 1} dz = \Gamma(x) \zeta(x),
\end{align}
where $\Gamma(x) = (x-1)!$ is the Euler gamma function, and $\zeta(x)$ is the Riemann zeta function, we get for the current
\begin{align}
\left( J_{x} \right)_{{\bm \Gamma}} = - \frac{5\zeta(5)}{9V \sqrt{3}\pi}  SD \left( \frac{T}{JS} \right)^{5} \nabla_{y}\chi 
= \frac{5\zeta(5)}{9V \sqrt{3}\pi} \frac{D}{J} \left( \frac{T}{JS} \right)^{4}\nabla_{y}T(\bf r) ,
\end{align} 
with an estimate $\zeta(5) \approx 1$.

\subsection{${\bf K}^{\prime}$ and ${\bf K}$ points}
Let us study the spectrum close to ${\bf K}^\prime = \left(0, \frac{4\pi}{3} \right)$
\begin{align}
&
\left( \Delta_{\bf k}\right)_{{\bf K}^\prime} \approx \Delta \left[ -3\sqrt{3} + \frac{3\sqrt{3}}{4}k^2 \right]
\\
&
\left( \gamma_{\bf k}\right)_{{\bf K}^\prime} \approx -\frac{\sqrt{3}}{2} (k_{y}+ik_{x}),
\\
&
\left( E_{\pm \bf k}/JS\right)_{\bf K^{\prime}} \approx 3 \mp 3\sqrt{3} \Delta - \frac{1}{8}k^2 \pm \Delta \frac{3\sqrt{3}}{4}k^2.
\end{align}
The Berry curvature for the $\mathrm{\Rmnum{1}}$ block at ${\bf K}^\prime$ point is
\begin{align}
\left[ \Omega^{[\mathrm{O}]}_{xy}({\bf k}) \right]_{{\bf K}^\prime}=  \frac{9}{8\left(\sqrt{9 - \frac{3}{4}k^2} \right)^{3}} \approx \frac{1}{24}
\end{align}

At ${\bf K} = \left(0, -\frac{4\pi}{3} \right)$ point we expand as
\begin{align}
&
\left( \Delta_{\bf k}\right)_{\bf K} \approx - \Delta \left[ -3\sqrt{3} + \frac{3\sqrt{3}}{4}k^2 \right]
\\
&
\left( \gamma_{\bf k}\right)_{\bf K} \approx \frac{\sqrt{3}}{2} (k_{y}-ik_{x}),
\\
&
\left( E_{\pm \bf k}/JS\right)_{\bf K} \approx 3 \pm 3\sqrt{3} \Delta - \frac{1}{8}k^2 \mp \Delta\frac{3\sqrt{3}}{4}k^2 .
\end{align}
The Berry curvature for the $\mathrm{\Rmnum{1}}$ block at ${\bf K}$ point is
\begin{align}
\left[ \Omega^{[\mathrm{O}]}_{xy}({\bf k}) \right]_{\bf K} =  -\frac{9}{8\left( \sqrt{9 - \frac{3}{4}k^2} \right)^{3}} \approx -\frac{1}{24}.
\end{align}
The Berry curvature for the $\mathrm{\Rmnum{2}}$ block is the same of the that of the $\mathrm{\Rmnum{1}}$ block. It can be deduced that the current is 
\begin{align}
\left( J_{x} \right)_{\bf K} =  \frac{4}{V} \sum_{{\bf k}} \left[ \Omega^{[\mathrm{O}]}_{xy} ({\bf k}) \right]_{\bf K} 
\left\{ c_{1}\left[ (E_{ -{\bf k}}  )_{\bf K}\right] - c_{1}\left[ ( E_{\bf k} )_{\bf K}\right] \right\}
\nabla_{y}\chi ,
\end{align}
assuming a small DMI, i.e. $\Delta = \frac{D}{J} < 1$, and expanding in $\Delta$, we get for the current
\begin{align}
\left(J_{x} \right)_{{\bf K}} = - \frac{3\sqrt{3}\Lambda^2}{8V\pi} SD \left(\frac{3JS}{T}e^{-\frac{3JS}{T}} \right)\nabla_{y}\chi
=  \frac{9\sqrt{3}\Lambda^2}{8V\pi} \frac{D}{J} \left( \frac{JS}{T}\right)^2 e^{-\frac{3JS}{T}} \nabla_{y}T(\bf r),
\end{align} 
where we introduced a high limit cut-off $\Lambda \sim 1$ on $k$, such that $\sum_{{\bf k}} \approx \frac{\Lambda^2}{4\pi}$ and we summed over all ${\bf K}$ points.

\begin{figure} \centerline{\includegraphics[clip,width=1\columnwidth]{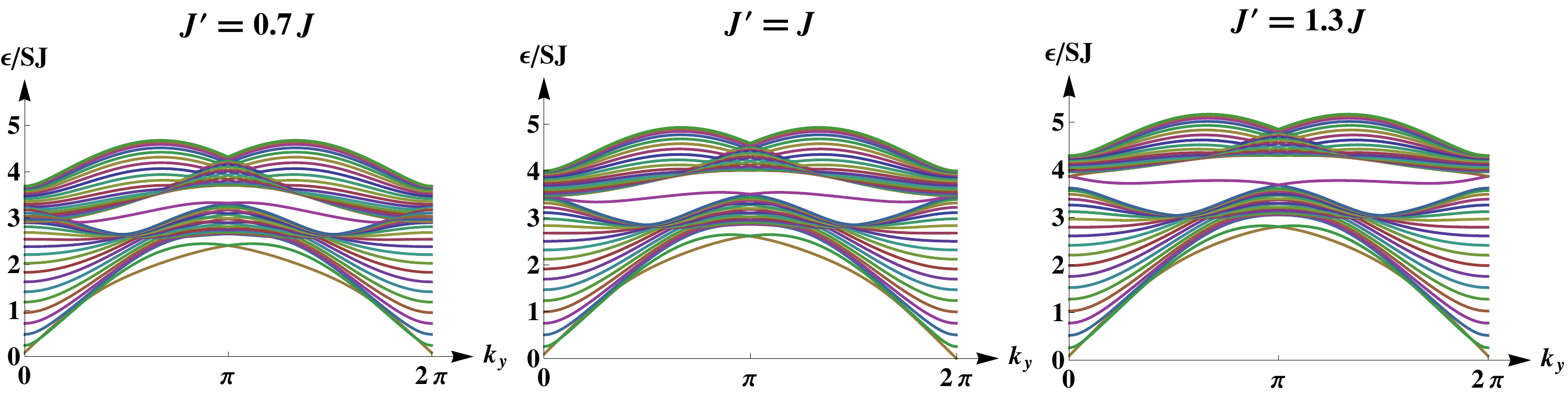}}

\protect\caption{(Color online)Evolution of the high energy edge states with increasing the interlayer coupling between the layers $J^{\prime}$. DMI strenght is $D=0.2J$.   }

\label{fig:Fig3}  

\end{figure}

\section{B model}
We study an antiferromagnet on a double layer honeycomb lattice. The Hamiltonian of the first block is given as 
\begin{align}
H =  JS \left[ \begin{array}{cccc}
\lambda + \Delta_{\bf k} & -\gamma_{\bf k} & 0 & t \\
-\gamma_{-\bf k} & \lambda -  \Delta_{\bf k} & t & 0 \\
0 & t & \lambda - \Delta_{\bf k} & - \gamma_{-\bf k} \\
t & 0 & -\gamma_{\bf k} & \lambda + \Delta_{\bf k} \\
\end{array}
\right],
\end{align}
and is described by a $\Psi_{\bf k} = ( a_{1{\bf k}},b_{1,-{\bf k}}^{\dag},b_{2{\bf k}},a^{\dag}_{2,-{\bf k}})^{\mathrm{T}}$ spinor, and we have defined $\lambda = 3+t$, where $t= J^{\prime}/J$.
The spectrum is derived 
\begin{align}
E_{\bf k \pm }^{2}/(JS)^2  = \lambda^2 - \vert \gamma_{\bf k} \vert^2 + \Delta_{\bf k}^{2} -T^2 \pm 2 \sqrt{\Delta_{\bf k}^{2}(\lambda^2 - \vert\gamma_{\bf k}\vert^2) + t^2\vert\gamma_{\bf k}\vert^2}.
\end{align}
In case of zero DMI, $\Delta = 0$, spectrum is
\begin{align}
E_{\bf k \pm }^2/(JS)^2 = \lambda^2 - t^2 \pm 2t\vert\gamma_{\bf k}\vert .
\end{align}
At energy $E_{\pm \bf k}/JS = \sqrt{\lambda^2 - t^2}$ there is a linear band touching at ${\bf K}^{\prime}$ and ${\bf K}$ points, where $\gamma_{\bf k} \approx \mp \frac{\sqrt{3}}{2} (\mp k_{y}- ik_{x})$ correspondingly. 
When the DMI is added, the ${\bf K}^{\prime}$ and ${\bf K}$ points get gapped. To see that, we set $\gamma_{\bf k} =0 $, we then get
\begin{align}
E_{\bf k \pm }/JS = \sqrt{ ( \lambda \pm \vert \Delta_{\bf k}\vert )^2 - t^2 }.
\end{align}
There is not much one can do analytically for this model.  
To study the evolution of high-energy edge states we plot the spectrum of a strip of double honeycomb antiferromagnet on Fig. [\ref{fig:Fig3}] .

\end{widetext}

\end{document}